\documentclass[ASNA,twocolumn]{USG}
\renewcommand{\orchid}[1]{\href{#1}{\textsf{\scriptsize ORCID}}}
\usepackage{mathtools,verbatim}
\ifdefined\XeTeXversion
  \usepackage{unicode-math}
  \renewcommand{\boldsymbol}[1]{\symbfit{#1}}
\fi
\graphicspath{{images/}}
\hypersetup{hypertexnames=false}
\newcommand{\bea}{\begin{eqnarray}}
\newcommand{\eea}{\end{eqnarray}}
\providecommand{\hl}[1]{#1}
\renewcommand{\hl}[1]{#1}
\newenvironment{greenblock}{\begingroup}{\endgroup}
\newcounter{lastmainfigure}

\startpage{1}

\makeatletter
\patchcmd{\printabstractpart}{\textbf{Correspondence:~}}{\textbf{E-mail:~}}{}{}
\renewcommand{\@@address}[2][]{\g@addto@macro\jmkaddress{\noindent #2\par}}
\let\@@articletype\relax
\let\oddhead@titlepage@info\@empty
\let\evenhead@titlepage@info\@empty
\patchcmd{\@maketitle}{\vskip24.36mm}{\vskip6mm}{}{}

\def\wiley@pagefoot#1{\vbox{\hsize=\textwidth\vskip3.5mm
  \hrule height .5pt\kern3pt\hbox to \textwidth{\pagenumfont #1}}}
\def\oddhead@headings@info{\wiley@pagefoot{\hfil\thepage}}
\def\evenhead@headings@info{\wiley@pagefoot{\thepage\hfil}}
\let\oddfoot@titlepage@info\oddhead@headings@info
\let\evenfoot@titlepage@info\evenhead@headings@info
\makeatother

\begin{document}
\title{Probing Topological Phases in a Strongly Correlated Ladder Model via entanglement}
\subtranstitle{trans-subtitle}

\makeatletter
\renewcommand{\@@address}[2][]{%
  \Newlabel{#1}{#1}%
  \g@addto@macro\jmkaddress{\noindent\textsuperscript{#1}\,#2\par}%
}
\makeatother

\author[1]{Aminul Hussain}[https://orcid.org/0009-0008-4640-1543]
\author[2,3]{Nisa Ara}
\author[3]{Rudranil Basu}
\author[1]{Sudeshna Sen}[https://orcid.org/0000-0003-2108-2183]

\address[1]{Department of Physics, Indian Institute of Technology, Dhanbad 826004, India}
\address[2]{Department of Theoretical Physics, Tata Institute of Fundamental Research,
Homi Bhabha Road, Mumbai 400005, India}
\address[3]{Department of Physics, Birla Institute of Technology and Science Pilani,
Zuarinagar, Goa 403726, India}

\corres{Rudranil Basu (\email{rudranilb@goa.bits-pilani.ac.in}), Sudeshna Sen (\email{sudeshna@iitism.ac.in})}

\transkeywords{Keyword1 | Keyword2 | Keyword3 | Keyword4 | Keyword5}

\abstract[ABSTRACT]{
\noindent
This work examines the interplay of band topology and interactions in a spinless two-leg ladder with $p$-wave-like inter-orbital hybridization.
In the non-interacting limit, the model hosts three topologically distinct insulating phases, denoted by $A$, $B$, and $C$.
Combining density-matrix renormalization group calculations with low-energy theories of the critical lines yields the interacting phase diagram for an inter-orbital density-density interaction.
The $A$-$C$ boundary contains two Dirac valleys and has central charge $c=2$ in the free theory. In the presence of interaction, the exact reflection symmetry pins the $A$-$C$ boundary to its free-theory location, while the critical theory evolves to $c=1$ and the edge entanglement entropy changes from $2\ln 2$ to $\ln 2$.
The $B$-$C$ boundary contains a single Dirac mode and shifts quadratically with interaction strength through second-order self-energy corrections involving virtual heavy modes, while retaining its $c=1$ critical structure.
Charge conservation and antiunitary many-body chiral symmetry give an interacting $\mathbb{Z}_4$ classification. This distinguishes the two interacting topological phases despite their identical lowest entanglement-spectrum degeneracy.
The boundary response is governed by the symmetry and operator content of its critical theory and the symmetry-protected topology of the adjoining phases.
}
\maketitle
\noindent

\section{Introduction}
\label{sec:intro}
Symmetry-protected topological (SPT) phases of matter, discovered a few decades ago, continue to attract significant attention because of their unconventional properties~\cite{Hasan_2010_RMP_Colloquium,Fu_Kane_Mele,Moore_Balents_2007,Roy2009,hasan2011review, Bansil_RMP_2016_Colloquium, Asboth2016}. Recent attention and awareness of quantum materials have further contributed to their intense research, opening new directions. Non-interacting topological insulators, with insulating bulk and conducting boundary states, are analyzed through electronic band theory~\cite{Bansil_RMP_2016_Colloquium}. There exist well-established {single particle band invariants} to identify the topological phases~\cite{Hasan_2010_RMP_Colloquium,Fu_Kane_Mele}. For example, there exist quantities like the quantized bulk Hall resistivity that can be connected to quantities of theoretical origin, called topological invariants, that can be expressed in terms of winding number ($\nu$)~\cite{Chang_RMP_2023_Colloquium} { and organised based on the Altland-Zirnbauer symmetry class of the Bloch Hamiltonian~\cite{Nersesyan_PRB_2020,Ghelli_2020_PRB_Topological,Mandal_PRB_2024_Topoogical,Li_PRB_2017_Topological,parida2025couplinginducedemergenttopology,zhou1,zhou2,Karmela_2018_PRB,Barbarino_PRA_2018_Topological,aghtouman2024dimerized}}.

For 1D systems with chiral symmetry (also called sublattice symmetry), the winding number $\nu$ acts as a topological invariant and is directly related to the Zak phase $\phi$ by the relation $\phi = \pi\nu$. When inversion symmetry is also present, the Zak phase becomes quantized to $0$ or $\pi$, corresponding to winding numbers $\nu = 0$ or $\nu = 1$~\cite{LEE2022winding}.
The physical properties related to these topological invariants are robust against perturbations as long as the insulating bulk gap \hl{remains open}~\cite{Hasan_2010_RMP_Colloquium, Qi_2011_RMP}. Topological invariants derived from bulk quantities form the bulk boundary correspondence, where properties of the bulk are associated with the conducting surface modes~\cite{Essin_2011_PRB_Bulk, Qi_2011_RMP}. Furthermore, different symmetry constraints on the Hamiltonian come into play so that different topological models can be classified based on these symmetries~\cite{Altland_1997_PRB_Nonstandard, Ryu_2010, Kitaev_2009_AIPConference_priodic, Chiu_RMP_2016_Classification}. { SPT phases provide a robust way of organizing gapped quantum matter beyond symmetry breaking.}

{ Particularly crucial is the understanding of topological phases in the presence of electronic correlations~\cite{senthil_review, Rachel_2018}.}
A hallmark of a topological quantum phase transition is the absence of a conventional Landau-Ginzburg type symmetry-breaking description~\cite{Senthil_2004_PRB_Quantum, Hasan_2010_RMP_Colloquium}; { the same holds true for certain electron-electron interaction driven transitions like the Mott metal insulator transition or the Kondo breakdown scenario.} On the other hand, quantum materials, including topology and electron-electron interactions, offer us an abundance of emergent quantum phases across { several realistic model systems}~\cite{Rachel_2018}. Even without topology, strongly correlated systems are known to exhibit emergent phases with unconventional orders in such non-perturbative regimes; examples include Mott insulators~\cite{mott_1968_Metal}, Wigner crystals~\cite{Wigner_1934_PR_On}, charge density wave ordering~\cite{Hatenyi_2020_PRR_Interaction}, domain wall crystal~\cite{Hu_PRB_1994_Domain}, and bond-density wave~\cite{Mazumdar_PRB_2000_Bond}. In many cases, the short or long range of the interactions matters and leads to different kinds of interaction-induced phase transitions~\cite{Kohn_1999_RMP_An_essay, Imada_1998_RMP_Metal, Defenu_RMP_2023_Long}. Quantum phase transitions are ubiquitous in these systems, resulting from non-perturbative and emergent physics prevalent in these settings~\cite{Rachel_2018, Kohn_1999_RMP_An_essay, Imada_1998_RMP_Metal}. 

Although a relatively new topic, the effects of strong correlations in topological systems have been explored in various settings~\cite {Rachel_2018, Qi_2011_RMP,Dzero_2016_AR,Dzero_2010_PRL,senthil_review}. While some studies focus on strongly correlated models like the topological Kondo insulator~\cite{Dzero_2016_AR,Dzero_2010_PRL} or topological Mott systems~\cite{Imada_1998_RMP_Metal,Raghu_2008_PRL}, there exist several other works on interacting SSH-type models~\cite{Salvo_2024_PRB,Melo_2023_PRB_Topological,Yu_2020_PRB_Topological,hang2025topological, matveeva2023one}. Although Hubbard-type onsite interactions are natural choices for capturing strong Coulomb interactions; such interactions, {in higher dimensions}, generally open a Mott gap at an otherwise critical Fermi-liquid description, thereby destroying the topological character at sufficiently strong coupling.  Particularly, in one dimension, a Fermi liquid description is absent, and a one-dimensional strongly correlated system is described as a Tomonaga-Luttinger liquid~\cite{schonhammer2004luttinger}. Other effects of interactions have recently been observed through the inclusion of interactions spanning over neighbouring sites. In general, the studies of interacting SSH-type models conclude that the non-interacting topological phase remains non-trivial even in the presence of interactions and undergoes a topological to charge density wave phase transition at some critical interaction strength~\cite{Salvo_2024_PRB, Melo_2023_PRB_Topological, Zhou_2023_PRB, Zhou_2025_PRB_Interaction}. Formation of anti-ferromagnetic order within the topological and trivial phase has also been reported recently for interaction spanning neighbouring sites~\cite{hang2025topological}. { But what happens when other type of scenarios like Kondo breakdown could be prevalent in the microscopic model? Indeed, in recent times, the interplay of Kondo physics and band topology in classic Kondo insulator systems, has led to the emergence of the field Topological Kondo Insulators~\cite{Sen_PRR_2020_Fragility,Baruselli_2015_PRB_Kondo,Baruselli_2015_PRL_Distinct, Alexandrov_2014_PRB_End}. Indeed one-dimensional models depicting conventional Kondo physics in topological settings has been explored and stable Haldane like phases have been identified including emergent energy scales and Kondo screening effect of the topological boundary state at play. These results naturally motivate how a generic, non-Kondo correlation would affect the pre-existing topology of the model. Our microscopic model is designed such that it can address the latter question.}

Non-interacting, topological systems can generally be easily understood using conventional free theory. Certain interacting topological systems in 2D and 3D can also be understood in terms of non-interacting electrons using a Fermi liquid description,~\cite{Wen_2012_PRB_Symmetry, Ryu_2010, Lundgren_2015_PRB_Landau}. However, such descriptions may break down as we increase the interaction strength and explore non-perturbative regimes~\cite{Sagi_2015_PRL_Breakdown,fidkowski2011topological}. { Fidkowski and Kitaev showed that interactions can reduce free-fermion classifications by gapping boundary degrees of freedom without breaking the protecting symmetry~\cite{Fidkowski_2010_PRL_Entanglement,fidkowski2011topological}.  More generally, interacting one-dimensional SPT phases are classified by the projective or graded projective realization of the protecting symmetries on the entanglement or edge Hilbert space~\cite{pollmann2010entanglement,pollmann2012symmetry,turner2011topological}.  A rigorous fermionic formulation associates one-dimensional fermionic SPT phases with indices built from fermion parity, symmetry fractionalization, and graded stacking data~\cite{BourneOgata2021, Han_Lake_Lam_Verresen_You_2024}.  When spatial symmetries are relevant, crystalline SPT phases require a separate but related framework; the crystalline equivalence principle gives a systematic way to compare spatial and internal protecting symmetries~\cite{ThorngrenElse2018}.  Recent work on one-dimensional charge-conserving fermions has made this connection particularly explicit for chiral- and reflection-protected SSH-type phases, showing how interactions reorganize the free winding classification into many-body invariants determined by the microscopic symmetry structure~\cite{LeeShiozakiHsieh2024}.  {These developments motivate the {other} question addressed in this work: once interactions are introduced in a multi-boundary SSH-like ladder, what is the actual interacting SPT label of each gapped phase? An associated question would concern the stability of the phase boundaries themselves as we include interactions.}}

Our lattice geometry is motivated in part by spinless $p$-wave periodic-Anderson-type models, where dispersive $c$ orbitals and more localized $f$ orbitals are coupled by an odd-parity hybridization~\cite{zhong-2018,Alexandrov_2014_PRB_End}. We use this structure as a platform for studying how a local $c$-$f$ density interaction modifies an existing topological phase diagram. The same orbital and hybridization structure also provides a natural starting point for future higher-dimensional extensions, including approaches based on dynamical mean-field theory.

{On the other hand, the model we probe can also be thought of as a tailored, ladder SSH model, where the two ladders or rungs are coupled in a special way via diagonal and direction-dependent hybridization energies.} We discuss limits where this model can render simple analytical solutions, and we predict the respective phase diagram. Ladder models provide a pathway to understand how the topological properties transform as we gradually increase dimensions. Ladder models provide a quasi-one-dimensional setting with additional transverse connectivity, allowing several microscopic routes to topological band structure within a controlled geometry. In fact, there can now be different sources of topology in the system in addition to the staggered hopping in an individual SSH chain~\cite{Nersesyan_PRB_2020, Ghelli_2020_PRB_Topological, Mandal_PRB_2024_Topoogical, Li_PRB_2017_Topological, parida2025couplinginducedemergenttopology,zhou1,zhou2}. Studies show that the SSH ladder behaves like a Kitaev chain depending on the inter-chain coupling and may also reproduce the classic Hofstadter butterfly pattern~\cite{Karmela_2018_PRB, Barbarino_PRA_2018_Topological,aghtouman2024dimerized}.

{Regarding the question about the phase boundary stability, we derive effective low energy theories in the continuum limit. This analysis is based on the idea that a topological transition between two free-fermion phases can often be described by a Dirac mass changing sign.In an interacting one-dimensional system, the mass channel may be renormalized, shifted, or even removed from the low-energy theory, depending on which microscopic interaction is projected into the critical subspace. Accordingly, the response of a phase boundary is governed jointly by the interacting SPT structure of the adjacent gapped phases and by the symmetry and valley content of the critical theory. Continuum bosonization and Green-function methods provide the natural tools for this question: bosonization determines whether a projected four-fermion operator gaps a collective sector, while a zero-frequency Green-function or self-energy analysis tracks the interacting renormalization of the Dirac mass~\cite{Giamarchi2004,Essin_2011_PRB_Bulk,Manmana2012InteractingInvariants}. Apart from the question of phase boundary stability, this also motivates another question addressed below: given a fixed local interaction, why can one topological boundary remain pinned while another boundary in the same model shifts?

Even at the non-interacting level, the designed model shows three distinct phases. We characterize the non-interacting topological phases by calculating the winding numbers ($\nu$) and also confirm the transition using the entanglement entropy values for the respective phases. Using the density matrix renormalization group (DMRG) method, we investigate inter-rung nearest neighbour electron-electron interactions to understand the stability of the non-interacting phase boundaries. We characterize the interacting topological phases using the entanglement spectrum and entanglement entropy values. We additionally provide a spectral gap analysis to identify the phase transitions occurring in the system. Finally, we provide an explanation of the robustness of the spectral gap by evaluating the central charge defined within an effective conformal field theory. We finally derive the continuum mechanism behind the different boundary responses {to augment our numerical calculations}. Such an analysis gives an interaction-class-dependent criterion for whether a topological boundary is pinned, shifted, split, or destabilized as interactions are included.}

The latter part of the paper is organized as follows. In Section~\ref{sec:model}, we introduce the model and discuss its single-particle and many-body protecting symmetries. The non-interacting topological phase diagram and its characterization through edge entanglement are presented in Sections~\ref{sec:non-int} and \ref{edge_corr}, respectively. In Section~\ref{sec:results_int}, we analyze the interacting results and many-body chiral indices along with critical central charges. Section~\ref{sec:free_critical_content} develops the low-energy descriptions, while Section~\ref{sec:interaction_class_viewpoint} discusses how these behaviors depend on the interaction class. Finally, we summarize our conclusions in Section~\ref{sec:conclusions}.

\section{Model and Symmetries}
% \vspace{-0.2 cm}
\label{sec:model}

\begin{figure}[htbp]
    \centering
    \includegraphics[width=1\columnwidth]{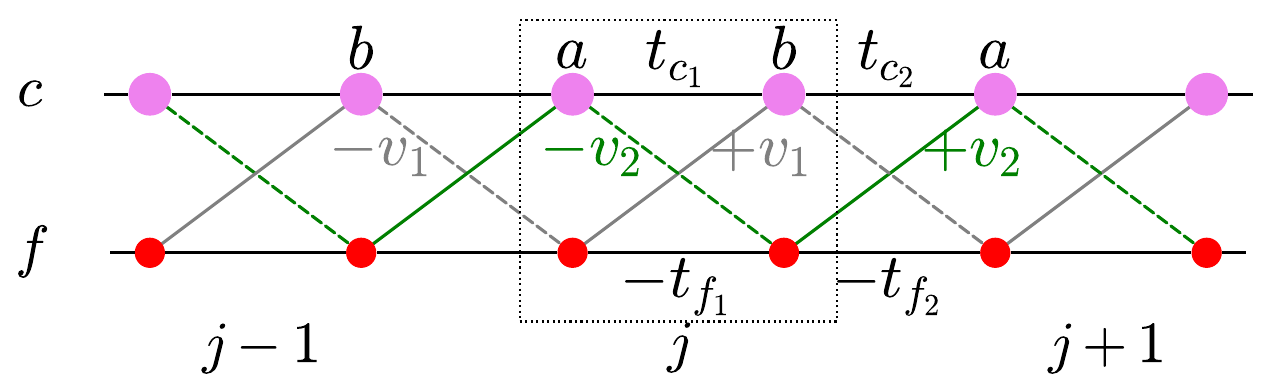}
    \caption{Schematic representation of the one-dimensional ladder model investigated in this work. Each unit cell (illustrated within a dotted square) consists of two sub-lattices, indicated by $a$ and $b$, the top (bottom) rung of each sub-lattice comprises the $c$ ($f$)-orbitals. The parameters: $t_{c_1}(t_{f_1})$ and $t_{c_2}(t_{f_2})$ represent the intra-lattice and inter-lattice hopping parameters for the $c\,(f)$ orbital, respectively; the orbitals hybridize via a $p$-wave-like hybridization energy given by $\pm v_{1}$ (between $c_b$ and $f_a$), and $\pm v_{2}$ (between $c_a$ and $f_b$).}
    \label{fig:model}
\end{figure}

In this section, we discuss the model Hamiltonian studied in this work. We also discuss the symmetries of the non-interacting system and discuss analytically tractable limits. 
The non-interacting, tight-binding Hamiltonian $H_0$ is shown in Figure~\ref{fig:model} and can be written as: 
\begin{align} 
    H_0 &= \nonumber \sum_{j} \left[ \left(t_{c_1} c^{\dagger}_{aj} c_{bj}-t_{f_1} f^{\dagger}_{aj} f_{bj}\right) + H.c \right] \\ \nonumber
    &+\sum_{j}\left[ \left(t_{c_2} c_{bj}^{\dagger} c_{aj+1} -t_{f_2} f_{bj}^{\dagger} f_{aj+1}\right)+H.c.\right] \\ \nonumber
    &+\sum_{j}v_{1}\left[\left(c_{bj}^{\dagger}-c_{bj-1}^{\dagger}\right)f_{aj}+f_{aj}^{\dagger}\left(c_{bj}-c_{bj-1}\right)\right] \\
    &+\sum_{j} v_{2} \left[\left(c_{aj+1}^{\dagger}-c_{aj}^{\dagger}\right)f_{bj}+f_{bj}^{\dagger}\left(c_{aj+1}-c_{aj}\right)\right],
    \label{eq:H0}
\end{align}
where, $c_{aj}$ ($c_{bj}$) denotes the annihilation operator for the $c$ electrons on the $a$-th ($b$-th) sublattice of the $j$-th unit cell. Similar notation is used for the $f$ electrons. 
Here, \( t_{c_1} \) and \( t_{c_2} \) refer to the \hl{ intra-lattice} and inter-lattice hopping parameters for \( c \) orbitals, whereas \( t_{f_1} \) and \( t_{f_2} \) indicate the \hl{ intra-lattice and} inter-lattice hopping parameters for \( f \) orbitals \hl{ respectively}. The parameters \( v_1 \) and \( v_2 \) are associated with hybridization between the $f$ and the $c$ orbitals. Note that the hybridization term is given as: $f_{aj}$ hybridizes with $c_{bj-1}$ and $c_{bj}$ with hybridization energy $-v_1$ and $+v_1$, respectively. Similarly, $f_{bj}$ hybridizes with $c_{aj}$ and $c_{aj+1}$ with hybridization energy $-v_2$ and $+v_2$, respectively. In other words, the hybridization between the two rungs (consisting of $c$ and $f$-orbitals) is represented via a direction-dependent hybridization energy. This direction dependence of the hybridization term is a crucial ingredient to realize topology in heavy fermion systems~\cite{Alexandrov_2014_PRB_End, ahamed2018rare}.

\begin{comment}
In order to explore the interplay of strong correlation and topology in the above model, we include a density-density interaction ( $H_{int}$) between the $c$ and $f$ orbitals, in Equation~\eqref{eq:H0}, where, 

\begin{align}
    {H_{\rm int} ={\pm} V_{NN}\sum_{j} \left(n_{c_a,j} n_{f_a,j}+n_{c_b,j} n_{f_b,j} \right),}
    %\quad n_{\alpha,j}=\alpha_j^\dagger\alpha_j
    \label{eqn:H_int1}
\end{align}
{and, $ n_{\alpha,j}=\alpha_j^\dagger\alpha_j$.}
\end{comment}
In order to explore the interplay of correlations and topology, we add a
local density-density interaction between the $c$ and $f$ orbitals on
each sublattice,
\begin{align}
H_{\rm int}
&= V_{NN}\sum_j \Bigl(\notag\\
&\quad n_{c_a,j}n_{f_a,j}\notag\\
&\quad +n_{c_b,j}n_{f_b,j}\Bigr),\notag\\
&\quad n_{\alpha,j}=\alpha_j^\dagger\alpha_j .
\label{eqn:H_int1}
\end{align}
Here $V_{NN}>0$ denotes repulsion and $V_{NN}<0$ attraction. Thus the
sign is included in the real parameter $V_{NN}$ itself.

We now discuss the non-interacting limit of the above model in more detail. The bulk spectrum of the current model can be obtained with the Bloch Hamiltonian, obtained by considering periodic boundary conditions (PBC) on a $N$ site lattice, given by, $c_{x{N+1}}^\dagger=c_{x1}^\dagger$, and, $f_{x{N+1}}^\dagger=f_{x1}^\dagger$, where $x$ denotes the sublattice index (a and b). Alternatively, one can consider the bulk to be the long central region of an open ladder. We represent the Bloch Hamiltonian in the chiral basis, $\Psi^{\dagger}_k=\left(c^{\dagger}_{ak},\, f^{\dagger}_{ak},\, c^{\dagger}_{bk},\, f^{\dagger}_{bk}\right)$, is given by, 

\begin{align}
    \mathcal{H}_{k} &= 
    \begin{pmatrix}
        0 & h(k) \\
        h(k)^{\dagger} & 0
    \end{pmatrix},
    \label{eq:chiral_basis_hamiltonian}
\end{align} 
where, 
\begin{align}
    h(k) &= \begin{pmatrix}
        t_{c_1}+t_{c_2}e^{-i k} & (e^{-i k}-1)v_{2} \\
        (1-e^{-i k})v_{1} &-(t_{f_1}+t_{f_2}e^{-i k})
    \end{pmatrix}. \label{eq:h(k)}
\end{align}

Rewriting $\mathcal{H}_{k}$ in terms of the Pauli matrices, we obtain, 
\begin{align}
    \mathcal{H}_k &= \Biggl\{ \left[ t_{cf1} + t_{cf2} \cos k \right] \sigma_x + t_{cf2} \sin k \, \sigma_y \Biggr\} \otimes \tau_0 \nonumber \\
    &\quad + \Biggl\{ \left[ t_{cf1}^{'} + t_{cf2}^{'} \cos k \right] \sigma_x + t_{cf2}^{'} \sin k \, \sigma_y \Biggr\} \otimes \tau_z
    \nonumber \\
    &\quad + v_{12} \Bigl[ (1-\cos k)\, \sigma_x - \sin k\, \sigma_y \Bigr] \otimes \tau_x 
    \nonumber \\
    &\quad + v_{12}^{'} \Bigl[ (1-\cos k)\, \sigma_y + \sin k\, \sigma_x \Bigr] \otimes \tau_y,
    \label{eq: pauli_decomposed_hamiltonian}
\end{align}
where, $(t_{c_1}-t_{f_1})/2 = t_{cf1},\, (t_{c_2}-t_{f_2})/2 = t_{cf2},\,(t_{c_1}+t_{f_1})/2 = t_{cf1}^{'},\,(t_{c_2}+t_{f_2})/2 = t_{cf2}^{'},\,(v_{1}-v_{2})/2 = v_{12}$ and $(v_{1}+v_{2})/2 = v_{12}^{'}$; the $\sigma$ matrices are written in the sublattice basis and the $\tau$ matrices project on to the respective orbitals. 

On the single-particle Hilbert-space, the sublattice operator $\sigma_z \otimes \mathbb{I}$, anti-commutes with the Bloch Hamiltonian $\mathcal{H}_k$, leading to the block off-diagonal representation in Equation~\eqref{eq:chiral_basis_hamiltonian}. { Because all hopping
amplitudes are real, the free Hamiltonian may also be organized as a BDI band problem and carries an integer winding number \cite{Altland_1997_PRB_Nonstandard,Ryu_2010, Kitaev_2009_AIPConference_priodic}. 

The interacting classification instead uses charge conservation and the antiunitary many-body chiral operation $\mathcal S=\mathcal C\mathcal T$ defined below. The subgroup used to protect the interacting phases is
$U(1)_Q\times\mathbb Z_2^{\mathcal S}$, with the antiunitary grading of
$\mathcal S$.}

We now briefly discuss some special limits of this model. The most trivial limit is the case when $v_1=v_2=0$, and the two chains of the ladder are decoupled, with the $c$-orbital and the $f$-orbital ladder taking the form of the SSH model, with two zero-energy edge modes in the topological phase. Further, if we now consider a limit where, $v_{12}^{'} = 0$ (i.e.,$v_1=-v_2$), and, $t_{c_1} + t_{c_2} =-(t_{f_1} + t_{f_2})$ provided $(v_{1},\, t_{c_2}+t_{f_2} \neq 0)$, we find an operator $P = \mathbb{I} \otimes \tau_z$ commutes with the Hamiltonian kernel. Hence, we can express the kernel in the eigenbasis of $P$, making it block diagonal. Consequently, with this choice of parameters, the model breaks down into two decoupled effective SSH models, $H_L$ and $H_R$, with corresponding effective hopping parameters:
\begin{align}
    u_\chi &= t_{cf1} \mp sR, &\qquad v_\chi &= t_{cf2} \pm sR. \quad \chi=L,R
    %\nonumber
    \label{eq:effective_hopping}
\end{align}
Here, $R=\sqrt{(t'_{cf2})^2+v_{12}^2}$ is non-negative, and $s=\operatorname{sgn}(t'_{cf2})$ encodes the original sign of $t'_{cf2}$. And hence the topological invariant of the whole system is $\mathbb{Z}_L \oplus \mathbb{Z}_R$; i.e., the direct sum of the topological invariants of the decoupled chains. Topological criteria for these decoupled chains are like the usual ones $|u_\chi|<|v_\chi|$. 
In Figure~\ref{fig:map_phase}, hatched regions with right tilted lines (/) and left tilted lines (\textbackslash) represent the parameter space for which $H_L$ and $H_R$ are in the topological phase, respectively. Clearly, in this limit, there could be a scenario where both SSH blocks are topological, thereby leading to four zero-energy topological edge modes for $\mathcal{H}_k$. We discuss this limit in more detail in Section~\ref{sec:non-int} and derive the condition, following~\cite{zou2015hidden}, in Appendix~\ref{app:decoupled_SSH}.

{
For the symmetry classification, let $L_{\rm uc}$ denote the number of
unit cells. The ladder then contains $4L_{\rm uc}$ fermionic modes. and is half filled at $N=2L_{\rm uc}$. It is useful to introduce the centered interaction
\begin{align}
H_{\rm int}^{\rm cen}
&=
V_{NN}\sum_j
\left(n_{c_a,j}-\frac12\right)
\left(n_{f_a,j}-\frac12\right)
\nonumber\\
&\quad+
V_{NN}\sum_j
\left(n_{c_b,j}-\frac12\right)
\left(n_{f_b,j}-\frac12\right).
\label{eqn:H_int_centered}
\end{align}
Expanding this expression gives
\begin{equation}
H_{\rm int}^{\rm cen}
=
H_{\rm int}
-\frac{V_{NN}}{2}N
+\frac{V_{NN}}{2}L_{\rm uc}.
\label{eq:centered_uncentered_relation}
\end{equation}
Consequently, in the fixed half-filled sector,
\begin{equation}
H_{\rm int}^{\rm cen}-H_{\rm int}
=
-\frac{V_{NN}}{2}L_{\rm uc},
\label{eq:centered_half_filling}
\end{equation}
so the centered and uncentered interactions have identical eigenstates
and differ only by an additive constant.

We now define the antiunitary many-body chiral operation $\mathcal S$ by \begin{align}
\mathcal S i\mathcal S^{-1}&=-i,
\nonumber\\
\mathcal S c_{a,j}\mathcal S^{-1}&=c_{a,j}^\dagger,
&
\mathcal S f_{a,j}\mathcal S^{-1}&=f_{a,j}^\dagger,
\nonumber\\
\mathcal S c_{b,j}\mathcal S^{-1}&=-c_{b,j}^\dagger,
&
\mathcal S f_{b,j}\mathcal S^{-1}&=-f_{b,j}^\dagger .
\label{eq:manybody_chiral_action}
\end{align}
It is evident from the above relations that $\mathcal S^2=1$. Acting on the
densities and total number,
\begin{equation}
\mathcal S n_{\alpha,j}\mathcal S^{-1}=1-n_{\alpha,j},
\qquad
\mathcal S N\mathcal S^{-1}=4L_{\rm uc}-N.
\end{equation}
The charge, relative to half-filling
\begin{equation}
Q=N-2L_{\rm uc},
\end{equation}
transforms as
\begin{equation}
\mathcal S Q\mathcal S^{-1}=-Q.
\end{equation}
Also, the corresponding unitary finite transformation generated by this charge is invariant:
\begin{equation}
\mathcal S e^{i\theta Q}\mathcal S^{-1}
=
e^{i\theta Q}.
\label{eq:S_commutes_U1}
\end{equation}
The symmetry used below is therefore
\begin{equation}
G_f=U(1)_Q\times\mathbb Z_2^{\mathcal S},
\label{eq:Gf_micro}
\end{equation}
where $\mathcal S$ is antiunitary and
$e^{i\pi Q}=(-1)^N$ is fermion parity.

For completeness, the direct invariance can be checked as follows. Let $s_\alpha=+1$ for an $a$-sublattice mode and $s_\alpha=-1$ for a $b$-sublattice mode. For two distinct modes,
\begin{equation}
\mathcal S\,
\alpha_i^\dagger\beta_j\,
\mathcal S^{-1}
=
s_\alpha s_\beta\alpha_i\beta_j^\dagger
=
-s_\alpha s_\beta\beta_j^\dagger\alpha_i.
\label{eq:S_bilinear_general}
\end{equation}
Every quadratic term in Eq.~\eqref{eq:H0} connects opposite sublattices, so $s_\alpha s_\beta=-1$ and each bilinear is mapped to its Hermitian conjugate. For example,
\begin{align}
\mathcal S c_{a,j}^\dagger c_{b,j}\mathcal S^{-1}
&=c_{b,j}^\dagger c_{a,j},
&
\mathcal S f_{a,j}^\dagger f_{b,j}\mathcal S^{-1}
&=f_{b,j}^\dagger f_{a,j},
\nonumber\\
\mathcal S c_{b,j}^\dagger c_{a,j+1}\mathcal S^{-1}
&=c_{a,j+1}^\dagger c_{b,j},
&
\mathcal S f_{b,j}^\dagger f_{a,j+1}\mathcal S^{-1}
&=f_{a,j+1}^\dagger f_{b,j}.
\end{align}
The $p$-wave hybridization differences are mapped to their Hermitian conjugates in the same way. Since all coefficients in the real-space Hamiltonian are real, $H_0$ is invariant under $\mathcal S$. Furthermore, every centered density changes sign, so each product in Eq.~\eqref{eqn:H_int_centered} is invariant. Thus
\begin{equation}
[\mathcal S,H_0+H_{\rm int}^{\rm cen}]=0
\end{equation}
at half filling.}

We focus on the interaction in Eq.~\eqref{eqn:H_int1} because it is the minimal local interaction coupling the $c$ and $f$ orbitals, directly competes with the inter-orbital hybridization $v_1,v_2$, and preserves both charge conservation and the antiunitary many-body chiral symmetry $\mathcal S$ at half filling. The spinless Hilbert space also permits high-accuracy DMRG calculations over the parameter ranges considered below. This provides a controlled setting for isolating how symmetry-preserving inter-orbital correlations reshape the topological phase diagram.

Before moving on to Section~\ref{sec:non-int}, we refer to a connection between the model studied in this work and the model of a topological Kondo insulator~\cite{Alexandrov_2014_PRB_End} that has gained some attention recently. Let us imagine a scenario where the $f$-electrons are strongly interacting, behaving as highly localized Wannier states or `local moment' spins. Then, the non-local, diagonal $c$-$f$ hybridization implies a $p$-wave type coupling. The lattice structure of certain heavy fermion materials, with strong spin-orbit coupling, indeed results in such non-local hybridization due to the crystal field environment~\cite{Alexandrov_2015_PRL, ahamed2018rare}. Its origin lies in the overlap between orbitals with different symmetries. In a specialized limit of the interaction strength (where a Hubbard Stratonovich transformation may be applied), this interacting $p$-wave Kondo insulator model maps onto an effective non-interacting model~\cite{Alexandrov_2014_PRB_End} that has the same structure as the Hamiltonian discussed in this work, albeit, $t_{c_1}=t_{c_2}$, $t_{f_1}=t_{f_2}$, $v_1=v_2$, \hl{ and redefining the unit cell}. In this limit, the sign difference in the diagonal hybridization is required to ensure the topological behaviour of the \textit{p}-wave Kondo insulator model\hl{, including the non-interacting case}. Thus, in this limit, the model supports exponentially localized edge states at each boundary with the topological invariant belonging to topological class D characterized by $\mathbb{Z}_2$~\cite{Alexandrov_2014_PRB_End}.

In this work, we consider the most general scenario of the non-interacting system, exploring the entire parameter space by systematically relaxing the above-discussed limits. As mentioned earlier, for interactions, we include a density-density interaction between the two orbitals residing on the two chains as shown in Equation~\eqref{eqn:H_int1}. In the following sections, we calculate and discuss the phase diagrams for the non-interacting system, followed by the interacting Hamiltonian.

\section{Phase Diagram}
\label{Sec:PhaseDiagram}
\subsection{Results: Non-interacting Topological Phases}
\label{sec:non-int}

Following the format prescribed in reference~\cite{matveeva2023one}, we can calculate the winding number for $V_{NN}=0$ using the formula 
$\nu = \frac{-i}{2 \pi}\int_{BZ} \partial_{k} i \phi(k) dk,$   
where $\phi(k)$ is the complex phase of the determinant of $h(k)$ Equation~\eqref{eq:h(k)}, and which can be defined as,
$\det h(k) = r(k) e^{i\phi(k)}$, $\phi (k) = \arctan \left[ \frac{Im \left[ \det h(k) 
 \right]}{Re\left[\det h(k)\right]}\right]$.

As mentioned in the previous section, the parameter regimes explored in our model include an interesting regime where the non-interacting Hamiltonian decouples into two independent SSH chains, with Hamiltonians indicated as $H_L$ and $H_R$ parametrized by effective hopping, Equation~\eqref{eq:effective_hopping}. These effective hopping parameters consequently determine the topological nature of individual chains independently, and hence the winding number of the entire system belongs to the direct sum of the winding numbers of individual decoupled chains(\(\mathbb{Z}_L \oplus \mathbb{Z}_R\)). Based on the topological nature of the individual chains ($H_L$ and $H_R$), the entire system's winding number may form in four ($0 \oplus 0$, $0 \oplus 1$, $1 \oplus 1$ and $1 \oplus 0$) distinct ways.

\begin{figure}[htbp]%[H]
    \centering
    \includegraphics[width=1\linewidth]{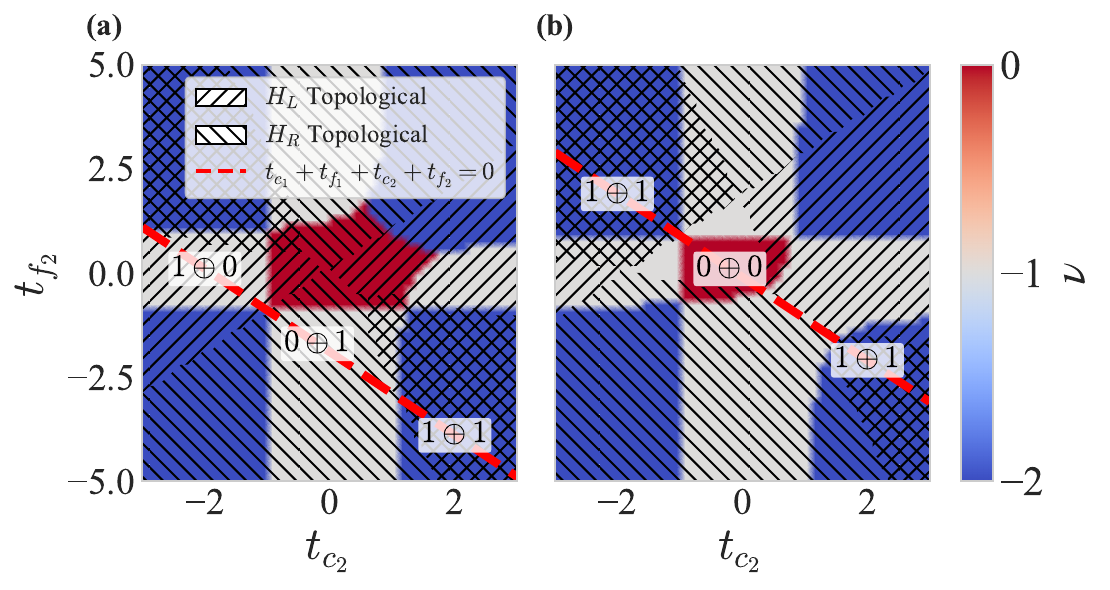}
    \caption{Solid colour regions represent the winding number $\nu$ calculated for $\mathcal{H}_k$ in the parameter space, $v_2/v_1=-1$. The colours blue, grey and red represent $\nu = 2, \, 1, \, 0$, respectively. We now calculate $\nu$ for two independent SSH chains, $H_L, \, H_R$ with hopping parameters defined in Equation~\eqref{eq:effective_hopping} irrespective of the condition $t_{c_1}+t_{f_1}+t_{c_2}+t_{f_2}=0$ but keeping $v_2/v_1=-1$. Hatched regions with right tilted lines (///) and left tilted lines (\textbackslash\textbackslash\textbackslash) represent the parameter space for which $H_L$ and $H_R$ are in the topological phase, respectively. The red dashed line represents the additional condition $t_{c_1}+t_{f_1}+t_{c_2}+t_{f_2}=0$ such that $\mathcal{H}_k=H_L\oplus H_R$. Hence, the winding number along the red dashed line can be written as (\(\mathbb{Z}_L \oplus \mathbb{Z}_R\)), and marked on the line. The left (right) plot is with $t_{c_1}=1$, $t_{f_1} = 0.9\,(-0.9)$.
    }
    \label{fig:map_phase}
\end{figure}

We get this regime by putting two constraints (i) $v_2/v_1=-1$ and (ii) $t_{c_1}+t_{c_2}+t_{f_1}+t_{f_2}=0$ in our model Hamiltonian $\mathcal{H}_k$, provided $(v_1,\, t_{c_2}+t_{f_2}\neq 0 )$. As a first step, we therefore explore this parameter regime and understand the respective phases. In Figure~\ref{fig:map_phase}, we demonstrate the phase diagram in the $t_{c_2}\,v/s\,t_{f_2}$ plane for $v_2/v_1=-1$, $t_{f_1}=0.9$ for Figure~\ref{fig:map_phase} (a) and $t_{f_1}=-0.9$ for Figure~\ref{fig:map_phase} (b). The different regions shown in Figure~\ref{fig:map_phase} can be understood as follows. Firstly, the solid coloured regions, purple, grey, and red, represent phases with total winding numbers of magnitude 2, 1, and 0, respectively. If we now solve for $H_L$ and $H_R$ (even in regions where $\mathcal{H}_k\ne H_L\oplus H_R$) independently in this parameter space, then we obtain the right(left) tilted hatched region when $H_L(H_R)$ is in the topologically non-trivial phase. Furthermore, the dashed red line represents the parameters where the total Hamiltonian decouples into $H_L$ and $H_R$, i.e., the regime where both the conditions, $(v_1=-v_2)$, and $( t_{c_1} + t_{c_2} =-(t_{f_1} + t_{f_2})$ are satisfied in the entire parameter space. We overlap these independently calculated phases on top of each other and observe where the respective (red dashed) line intersects with the hatched and solid coloured regions to identify whether the topological phases of the total Hamiltonian in the above regime, belong to (\(1 \oplus 0\), (\(0 \oplus 1\)), (\(1 \oplus 1\)) phase. In Figure~\ref{fig:map_phase} (a), with $t_{f_1}=0.9$, one can only identify topologically non-trivial phases with the above characterization. In Figure~\ref{fig:map_phase}(b), with $t_{f_1}=-0.9$, we can additionally identify a topologically trivial phase with the winding number (\(0 \oplus 0\)), both the decoupled chains are in trivial phase, represented by the intersection of the red-dashed line, the red solid region, and the unhatched region.
%\clearpage

\begin{figure}[htbp]%[H]
    \centering
    \includegraphics[width=1\linewidth]{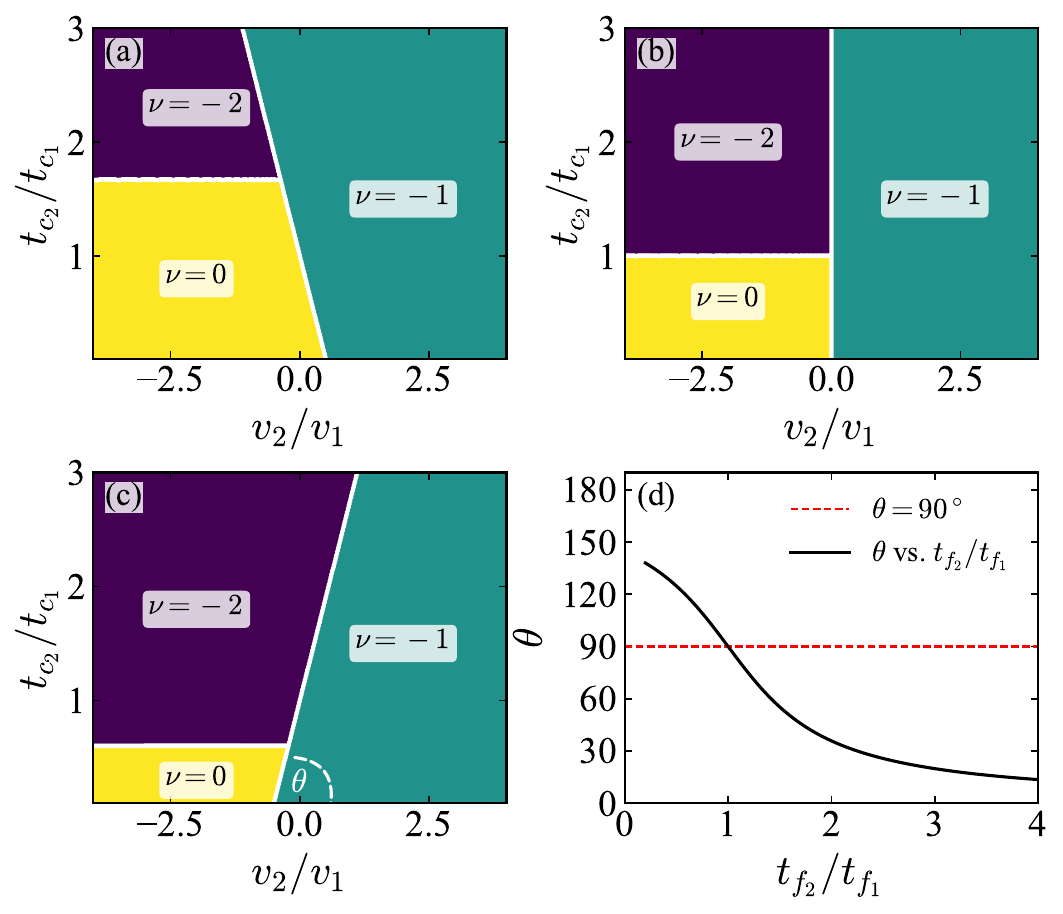}
    \caption{Phase diagram for the non-interacting ($V_{NN}=0$) case, in the $t_{c_2}/t_{c_1}$-$v_2/v_1$ plane, for different $t_{f_2}/t_{f_1}\approx 0.6, \, 1.0,\, 1.67$ in panels (a), (b), (c), respectively. The phase diagram consists of a trivial ($\nu=0$) and two different topological phases ($\nu=-2,\, -1$), characterized by different winding number ($\nu$) values as mentioned in the panels. Note that the slope of the phase separating line between phases with $\nu = -2, \, 0$ from the phase with $\nu = -1$ changes orientation depending on the value of $t_{f_2}/t_{f_1}$. This variation of the slope angle as a function of $t_{f_2}/t_{f_1}$ is shown in panel (d). The exact expression is shown in Equation~\eqref{eq: theta}}
    \label{fig:PhaseDiagram_nu}
\end{figure}

We now relax the above condition and calculate the winding number for the non-interacting system, keeping $t_{f_2}/t_{f_1}$ fixed and varying $t_{c_2}/t_{c_1}$ and $v_2/v_1$. In Figure~\ref{fig:PhaseDiagram_nu}, we plot the winding number $\nu$ in the $t_{c_2}/t_{c_1}$-$v_2/v_1$ plane, for different $t_{f_2}/t_{f_1}\approx 0.6, \, 1.0,\, 1.67$ as shown in panels (a), (b), (c), of Figure~\ref{fig:PhaseDiagram_nu}, respectively. Clearly, two different topologically non-trivial phases exist as indicated by the purple region ($\nu=-2$) and the green region ($\nu=-1$). As we fix $t_{c_2}/t_{c_1}$ somewhere in the purple region, with four degenerate zero energy edge states and increase $v_2/v_1$, we encounter a critical $v_2/v_1$ (say, $v_c$), where two of these degenerate states merge into the bulk. We end up with the green region, where $\nu=-1$. As shown in  Figure~\ref{fig:PhaseDiagram_nu}(b), $v_c$ remains constant (independent of $t_{c_2}/t_{c_1}$ when $t_{f_2}/t_{f_1}=1$. The critical phase boundary between the $\nu=-1$ and $\nu=-2$ phase and the $\nu=0$ and $\nu=-1$ phase however may vary as a function of $t_{c_2}/t_{c_1}$ and $v_2/v_1$ as we deviate from $t_{f_2}/t_{f_1}=1$. This is shown in panels (a) and (c) of Figure~\ref{fig:PhaseDiagram_nu} for two values of $t_{f_2}/t_{f_1}$, and the variation of the angle of this phase boundary with the $v_2/v_1$ axis is plotted in panel (d) of Figure~\ref{fig:PhaseDiagram_nu}. 
One can also evaluate this phase boundary behaviour through an exact expression derived in Appendix~\ref{app:exact_calc}. And the exact expression of the angle plotted in panel (d) is
\begin{align}
    \theta = \arctan{\left(\frac{4v_{1}^{2}}{(\frac{t_{f_2}}{t_{f_1}}-1)t_{f_1}t_{c_1}}\right)}.
    \label{eq: theta}
\end{align}

The Zak Phase or winding number undoubtedly works well for non-interacting systems. The same, however, may fail to provide information in an interacting system~\cite{Zhao_2023_PRL_Failure}. For interacting systems, the topological order parameters can be defined in terms of response functions. A general topological order parameter may also be defined in terms of the zero-frequency Green's functions~\cite{wang2012simplified}. 
In recent years, quantities based on mutual information theories have emerged in this category of behaving as order parameters for quantum phase transitions. In particular, analysis based on the entanglement spectrum or the von-Neumann entanglement entropy (EE) has indeed served as a useful tool to detect quantum phase transitions.

\subsection{Results: Edge Entanglement}\label{edge_corr}
In this section, we re-analyze the non-interacting phase diagram Figure~\ref{fig:PhaseDiagram_nu} in terms of the von-Neumann EE. In Section~\ref{sec:results_int}, we calculate the EE for an interacting system to derive the phase diagram for the interacting Hamiltonian.

\begin{figure}[htbp]
    \centering
    \includegraphics[width=0.9\linewidth]{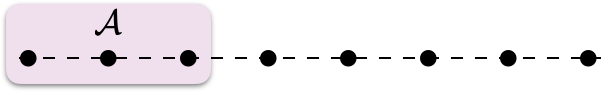}
    \caption{Cartoon of a finite-sized chain, with the subsystem $\mathcal{A}$ chosen such that it includes one of the edges.}
    \label{fig:EE_bipartition}
\end{figure}

For free fermionic systems in an SPT phase, the lower bound on sub-system von Neumann EE qualifies as a topological invariant equivalent to the winding number/ Zak phase described above~\cite{ryu2006entanglement}. To understand the non-trivial aspect of the quantized entanglement between the edge modes, let's first focus on the EE of a subregion $\mathcal{A}$ containing one of the edges of a finite chain, with the rest of the degrees of freedom in the chain, with open boundary conditions (OBC), as depicted in Figure~\ref{fig:EE_bipartition}. 

For such a theory, EE in the ground state $\rho = |\psi \rangle \langle \psi |$ for a subregion $\mathcal{A}$ is given by the Schmidt values or the entanglement spectrum obtained from the correlation functions~\cite{peschel2009reduced}, as:
\bea
    S_{\mathcal{A}} =-\sum_k \zeta_k\ln{\zeta_k}+(1-\zeta_k)\ln{(1-\zeta_k)}.
    \label{S_schmidt}
\eea
Here $\zeta_k$ are the eigenvalues of the correlation matrix supported on the region $\mathcal{A}$. 

For free theories, the Schmidt values, and hence $S_{\mathcal{A},\mathrm{OBC}}$ etc. Equation~\ref{S_schmidt} can be analytically computed using the two-point correlators, as such correlators are computed directly from the eigenvalues of the quadratic Hamiltonian's kernel. For example, if $V_{\alpha}$ are the eigenvectors of the real-space Hamiltonian kernel $\mathcal{H}$, such that the 2nd-quantized Hamiltonian is $H = \sum_{ij}\mathcal{H}_{ij} c^\dagger_i c_j$, then the correlation function in the half-filled ground state is given by $C_{ij} = \langle \psi | c^\dagger_i c_j | \psi \rangle = \sum_{\alpha}V_{\alpha}^{\dagger}{}_i V_{\alpha\,j}$. For a particle-hole symmetric Hamiltonian, whose ground state is filled with only valence band excitations, $\alpha$ in the sum runs over those eigenvectors, whose eigenvalues are negative.

For OBC, the half-filled ground state of a chain of $N$ sites in a trivially gapped phase takes the form:
\bea
    |\psi \rangle = \prod_{k=1}^{N} b^\dagger_k |0\rangle.
\eea

Here, $b_k$ are the negative energy modes, which are dispersive in the bulk. However, in a topological phase with zero-energy edge modes, the ground state is degenerate. For example, if there are two edge states and there are $N-2$ dispersive states, all the states:

\bea  &\prod_{k=1}^{N-2} b^\dagger_k |0\rangle, |e_1\rangle \otimes\prod_{k=1}^{N-2} b^\dagger_k |0\rangle, \nonumber \\ & |e_2\rangle \otimes\prod_{k=1}^{N-2} b^\dagger_k |0\rangle , |e_1\rangle \otimes |e_2\rangle \otimes\prod_{k=1}^{N-2} b^\dagger_k |0\rangle  \nonumber\eea 
in the thermodynamic limit are degenerate. In a finite-size chain, the energies of $|e_{1,2}\rangle$ are finite, equal and opposite, though exponentially suppressed to a power of $N$. Hence, if the energy of $|e_1\rangle$ goes to $0^-$ as $N \rightarrow \infty$, then $|e_1\rangle \otimes\prod_{k=1}^{N-2} b^\dagger_k |0\rangle$ is the ground state for finite $N$. It is interesting to point out that $|e_1 \rangle$ has non-zero support at the edges of the chain, with finite skin-depth penetrating the bulk~\cite{delplace2011zak}. Or, in other words, if $c_1$ and $c_N$ are the site local fermions sitting at the edges, the amplitudes $\langle 0| c^{\dagger}_1 |e_1 \rangle$ and $\langle 0|c^{\dagger}_N |e_1 \rangle$ give non-zero probabilities, whereas the support of the wavefunction decays exponentially as one gets deeper in bulk. Hence, the edge modes can render a Bell pair to the ground state, which reflects in the EE for a sub-system like $\mathcal{A}$ as per Figure~\ref{fig:EE_bipartition}.

The OBC and PBC bipartitions allow the boundary contribution to be
separated from the bulk entanglement. For a subsystem attached to a
physical end, the OBC geometry contains one bulk entanglement cut,
whereas the corresponding PBC bipartition contains two equivalent bulk
cuts. Thus one half of the PBC entropy gives the bulk contribution
associated with a single cut, and the edge contribution is
\begin{equation}
S_{\mathrm{edge}}
=
S_{\mathcal A,\mathrm{OBC}}
-\frac{1}{2}S_{\mathcal A,\mathrm{PBC}}.
\label{edge_def}
\end{equation}
This construction isolates the contribution associated with the
physical boundary~\cite{Calabrese:2004eu}.

\begin{figure}[htbp]
    \centering
    \includegraphics[width=0.9\linewidth]{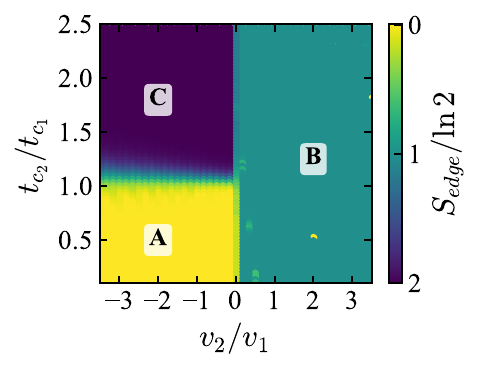}
    \caption{Phase diagram of the non-interacting system, in the $t_{c_2}/t_{c_1}$-$v_2/v_1$ plane, for $t_{f_2}/t_{f_1}=1$, reproduced in terms of the edge entanglement entropy, $S_{edge}/\ln(2)$. The phases are in excellent agreement with that derived using $\nu$ (Figure~\ref{fig:PhaseDiagram_nu}(b)). The quantity, $S_{edge}/\ln(2)$, thus serves as an excellent order parameter for identifying topological phase transitions in the non-interacting system. Phases A, B, and C are labelled based on the $S_{edge}/\ln{2}$ values, which are equal to 0, 1, and 2, respectively.}
    \label{fig:Non-int_S_edge_by_ln2}
\end{figure}

In Figure~\ref{fig:Non-int_S_edge_by_ln2}, we plot the phase diagram using $S_{\mathrm{edge}}$ as the order parameter, as defined in Equation~\ref{edge_def}. For the contributions of the OBC and PBC parts in this, we use the eigenvalues from the subsystem correlation matrix, as given in Equation~\ref{S_schmidt}. The Hamiltonian parameters used to generate this figure are exactly the same as those used in Figure~\ref{fig:PhaseDiagram_nu}(b).

The conclusion from the last section benchmarks edge-entanglement to faithfully reproduce the winding number calculations for the topological phases of the Hamiltonian. This is important in the light of our forthcoming analysis of the fate of these topological phases in the presence of an interaction. As winding numbers, which are functionals of the Bloch wavefunctions, are well defined only for free theories, we will use the entanglement spectrum to probe the phases in the interacting model. We conclude this section with a brief outline of the calculation of the entanglement spectrum used to distinguish the different phases in the interacting Hamiltonian.

%\clearpage
The entanglement spectrum is a classic tool for identifying a topological phase and serves as an unambiguous method to understand the topological structure irrespective of whether interactions are present or not. Considering the bipartitioning strategy discussed in Figure~\ref{fig:EE_bipartition}, we can compute the entanglement spectrum of a finite-sized open chain, where the part $\mathcal{A}$ is chosen to include one of the edges.
Since the reduced density matrix $\rho_A$ is a positive semi-definite Hermitian operator, we can express it in the form of the exponential of another Hermitian operator, which we refer to as the \textbf{entanglement Hamiltonian}, denoted as $H_E$. This relationship is represented as, $\rho_A = \frac{1}{Z} e^{-H_E}$. Here, $Z = \text{Tr}(e^{-H_E})$ ensures that $\text{Tr}(\rho_A) = 1$~\cite{fidkowski2011topological,peschel2009reduced, pollmann2010entanglement, pollmann2012symmetry, turner2011topological}. 

The eigenvalues of the entanglement Hamiltonian, represented as \(\{\xi_n\}\), possess a direct relationship with the eigenvalues $\{\lambda_n\}$ of the reduced density matrix. By diagonalizing both operators in the same basis, we derive the following relationship for each corresponding pair of eigenvalues: $\lambda_n = \frac{1}{Z} e^{-\xi_n}$.
This follows directly $\xi_n = -\log(\lambda_n) - \log(Z)$.
%\[\xi_n = -\log(\lambda_n) - \log(Z) \tag{1}\].
One can now readjust the position of the lowest eigenvalues of the entanglement Hamiltonian to zero by the introduction of an overall constant. This results in a more standardized expression, given by $\xi_n = -\log(\lambda_n)$. The entanglement spectrum denoted by the set of eigenvalues $\{\xi_n\}$ contains information about the topological content of the system. Li and Haldane proposed and demonstrated that the entanglement spectrum of the reduced density matrix of ground states in time-reversal-breaking topological phases reveals key information about their edge modes~\cite{Li_2008_PRL_Entanglement}. This proposal is also utilized and supported in several other works~\cite{Chandran_2011_PRB_Bulk, Turner_2010_PRB_Entanglement, Fidkowski_2010_PRL_Entanglement}.

\subsection{Results: Interacting Topological Phases}
\label{sec:results_int}
The interplay between topological non-triviality and strong correlation gives rise to interesting quantum phases. As a first step towards this in the model of our interest, we would need to quantify the effect of interaction on the phases otherwise distinguished by topological invariants in the free theory, which was described above. 

However, for interacting systems, the traditional approaches, like perturbative calculations and mean field approximations, are only suitable for weak interaction strengths~\cite{sirker2014boundary,junemann2017exploring}. Moreover, due to the inefficiency of mean field approaches to capture long-range correlations, they exhibit false or inaccurate critical points~\cite{kurita2016stabilization}.

We estimate the transition points from minima of the finite-size many-body gap calculated under PBC. A robust bulk transition requires the minimum to approach zero under finite-size scaling; at the sizes studied here, the position of the minimum provides an indication of the corresponding critical point. However, to characterize the phases, we need an order parameter. Multiple studies have successfully used edge entanglement entropy(EEE)~\cite{Wang_2015_PRB_Detecting, Ara_2024} as an order parameter to differentiate the topological phases. For this precise reason, we have benchmarked Figure~\ref {fig:Non-int_S_edge_by_ln2}, the same for the free part of our model.

All analyses in this section focus on the interaction in
Eq.~\eqref{eqn:H_int1}. Unless otherwise specified, we set
\[
t_{c_1}=1,\qquad v_1=-0.3,\qquad
t_{f_1}=t_{f_2}=0.3.
\]
Throughout the numerical sections, $L$ denotes the total number of
microscopic fermionic orbitals, equivalently the number of MPS sites,
so that
\begin{equation}
L=4L_{\rm uc}.
\end{equation}
The system size used for each calculation is stated together with the
corresponding result. In particular, we use $L=256$ for the edge
entanglement and $L=512$ for the number-density calculation. The
edge-entanglement calculation requires ground states under both OBC
and PBC; $L=256$ provides a useful balance between resolution of the
transition and the computational cost of the PBC calculation.

In this work, we perform the density-matrix renormalization group (DMRG) numerical method, which efficiently captures low-energy physics with strong correlations for one-dimensional systems~\cite{white1992density,Schollwock2011}. We mapped the ladder system to a pseudo-one-dimensional system. In order to benchmark the code for interacting systems, we employ a \hl{Gaussian}-MPS (GMPS) ansatz described in~\cite{fishman2015compression} to construct the ground-state for the ladder system for $V_{NN}=0$. We check for the convergence of ground-state energies and the bipartite von Neumann entanglement entropy by comparing it with the results obtained from the method of correlation functions in Equation~\ref{edge_corr}. We explored bond dimensions up to $\chi = 1500$ to ensure a $10^{-8}$ convergence for the ground-state energy and entanglement entropy. We then use this ground state obtained from the GMPS ansatz as an initial guess wavefunction for performing DMRG with the interactions.

\subsubsection{Edge Entanglement}
\label{sec:results_int_S_edge}

Motivated by the non-interacting phase diagram depicted in Figures~\ref{fig:Non-int_S_edge_by_ln2}, ~\ref{fig:PhaseDiagram_nu}(b), we focus on three different parameter regimes to identify how the non-interacting phase boundaries modify in the presence of nearest-neighbour $c$-$f$ density-density interactions, given by Equation~\eqref{eqn:H_int1}.
\begin{figure}[htbp]
    \centering
    \includegraphics[width=0.9\linewidth]{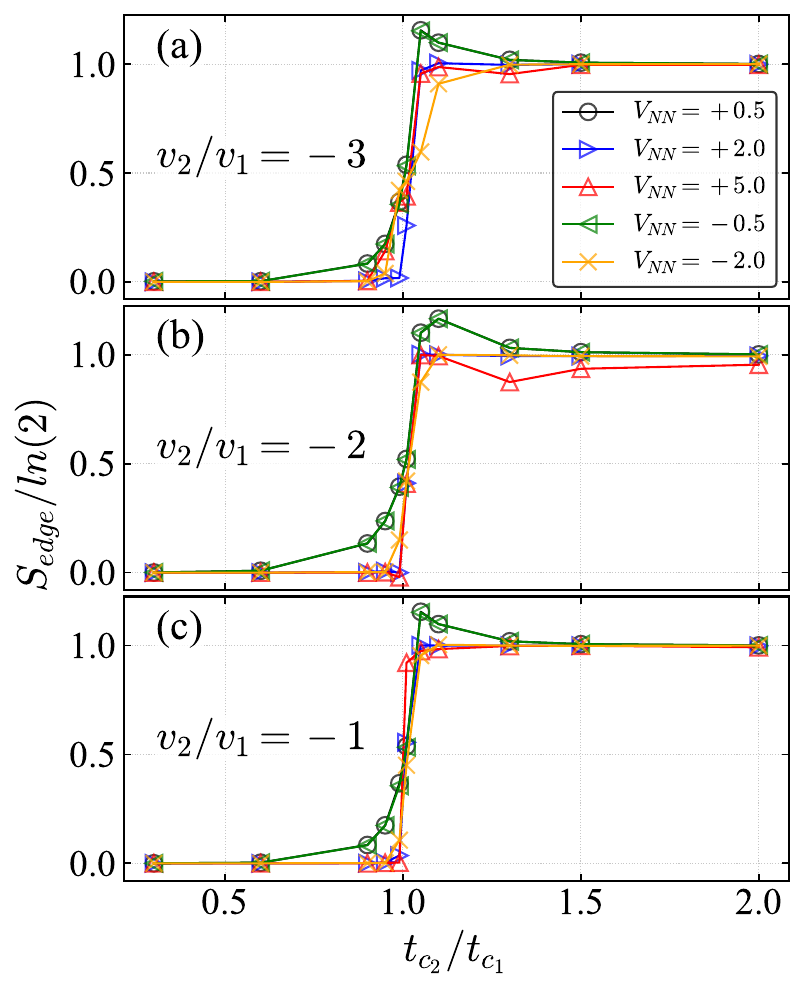}
    \caption{We plot $S_{edge}/\ln{2}$ as a function of $t_{c_2}/t_{c_1}$ for different repulsive and attractive interaction strengths, $V_{NN}$(see legends), are shown in sub-figure (a), (b) and (c) for $v_2/v_1=-3, \, -2,\, -1$, respectively. The lines connecting the symbols serve as a visual guide. The system size taken for this calculation is $L=256$. For the nonzero interaction strengths shown, the transition remains located at $t_{c_2}/t_{c_1}=1$. The $A$-$C$ boundary is therefore pinned within the numerical resolution.}
    \label{fig:EE_for_U}
\end{figure}
For all the interacting calculations, we fixed $t_{f_2}/t_{f_1}=1$ unless otherwise mentioned. Referring to Figure~\ref{fig:Non-int_S_edge_by_ln2}, we first focus on the left side of the $v_2/v_1=0$ boundary, i.e., the phases A and C. For this we calculate $S_{\mathrm{edge}}/\ln 2$ for $V_{NN}=\pm0.5, \pm2.0, +5.0 $, and plot them as a function of $t_{c_2}/t_{c_1}$ in Figure~\ref{fig:EE_for_U} for $v_2/v_1=-3,\, -2,\, -1$ in panels (a, b, c), respectively. 
For all these calculations, we used a system size of 256 sites. We observe that irrespective of the value of $v_2/v_1$, $S_{\mathrm{edge}}/\ln 2=0$ for all $t_{c_2}/t_{c_1}<1$ sharply crossing over to a value  $S_{\mathrm{edge}}/\ln 2\approx 1$ for all $t_{c_2}/t_{c_1}>1$. 
This holds true irrespective of the sign of $V_{NN}$. The crucial difference between the non-interacting and interacting results is that while in the former(Figure~\ref{fig:Non-int_S_edge_by_ln2}), $S_{\mathrm{edge}}/\ln 2$ transforms from a value of $0$ to $2$ while going from phase A to phase C, in the interacting case, this transition is marked by a change in the value of $S_{\mathrm{edge}}/\ln 2$ from $0$ to $1$. And hence there is a quantized jump of edge entanglement entropy from $2\ln{2}$ to $\ln{2}$ of non-interacting phase C when interaction is introduced. Nonetheless, the phase separation line continues to exist at $t_{c_2}/t_{c_1}=1$ and is robust even in the presence of such a type of nearest neighbour interaction. This is consistent with the behaviour of edge entanglement under the influence of interactions in simpler models with non-trivial topological phases~\cite{Wang_2015_PRB_Detecting, Ara_2024}. This is moreover supported by the fact that phase C  still has a density imbalance at the edges, whereas particle density remains uniform in phase A, even in the presence of strong correlation. The number density plots are shown in Appendix~\ref{app:number_density}.

\paragraph*{\textbf{Discussion for Edge Entanglement Result:}}

To understand the physics behind the jump of edge entanglement from $2\ln{2}$ to $\ln{2}$ in phase C of the phase diagram, Figure~\ref{fig:Non-int_S_edge_by_ln2}, we have done a simple perturbative calculation.
\paragraph*{Edge Hilbert Space:}
At zero energy, the relevant edge sector consists of one fermion in leg $c$ and one fermion in leg
$f$. Denoting the two edges by $p = 1, 2$, we define creation operators as:
\begin{equation}
 c_{1,A}^\dagger, \quad  c_{2,B}^\dagger, \quad
 f_{1,A}^\dagger, \quad  f_{2,B}^\dagger.
\end{equation}
Here we have assumed that the edge zero-modes are localized on specific sublattices: edge 1 on sublattice $A$, edge 2 on sublattice $B$. 

Now, the four edge basis states are (dropping the edge indices p)
\begin{align}
\nonumber
|1,1\rangle &= c_A^\dagger f_A^\dagger |0\rangle \quad
|2,2\rangle = c_B^\dagger f_B^\dagger |0\rangle \\  
|1,2\rangle &= c_A^\dagger f_B^\dagger |0\rangle \quad
|2,1\rangle = c_B^\dagger f_A^\dagger |0\rangle.
\end{align}
A general normalized edge state is
\begin{equation}
|\text{edge}\rangle = \sum_{p,q \in {1,2}} A_{p,q} |p,q\rangle, \qquad \sum_{p,q}|A_{p,q}|^2=1.
\end{equation}
Hence, in the non-interacting case, \textit{the edge state is 4-fold degenerate}. 

\paragraph*{Action of Interaction:}
The interaction at the edge is (dropped the edge indices p)
\begin{equation}
H_{\text{int}}^{(\text{edge})} = (V_{NN})(n_{c,A}n_{f,A}+n_{c,B}n_{f,B}).
\end{equation}
where $n_{x,y} = x^{\dagger}_{y}x_{y} \quad x \rightarrow \{c,f\}$ and $y \rightarrow \{A,B\} $
acting on each basis state explicitly:
\begin{align}
\nonumber
H_{\text{int}}^{(\text{edge})}|1,1\rangle 
&= V_{NN} |1,1\rangle, \quad H_{\text{int}}^{(\text{edge})}|1,2\rangle = 0,
\\
H_{\text{int}}^{(\text{edge})}|2,2\rangle &= V_{NN} |2,2\rangle, \quad H_{\text{int}}^{(\text{edge})}|2,1\rangle = 0.
\end{align}

For repulsive interaction, $V_{NN}>0$, the lower-energy subspace is spanned by $|1,2\rangle$ and $|2,1\rangle$, while $|1,1\rangle$ and $|2,2\rangle$ are shifted upward. For attractive interaction, $V_{NN}<0$, the ordering is reversed and the lower-energy subspace is spanned by $|1,1\rangle$ and $|2,2\rangle$. Thus either sign of a nonzero interaction splits the idealized four-dimensional edge manifold into two doublets. This edge-limit calculation accounts for the observed reduction of the lowest entanglement degeneracy. The interacting SPT label is determined independently from the many-body chiral edge algebra in Sec.~\ref{sec:manybody_spt_index}.

For a finite open chain, the exponentially localized edge modes form a
nearly degenerate low-energy subspace. A definite basis in this
subspace may be selected either through symmetric and antisymmetric
linear combinations of the edge modes or through an auxiliary
end-to-end hopping whose induced splitting remains well below the bulk
gap. This makes the edge contribution to the entanglement well defined
away from criticality.

In the interacting calculations, $S_{\rm edge}$ provides a direct
diagnostic along the $A$-$C$ cuts, where the transition remains at
$t_{c_2}/t_{c_1}=1$. For the boundaries involving phase $B$, we first
locate the interaction-dependent critical point from the many-body gap
and then characterize the adjacent gapped regions through the
entanglement spectrum. The gap and entanglement diagnostics therefore
play complementary roles in constructing the interacting phase
diagram.

\subsubsection{Spectral Gap Analysis}
\label{sec:spectral_gap}
We will determine the interacting phase boundaries from the many-body spectral gap under PBC and subsequently use the entanglement spectrum to characterize the resulting gapped regions. In this section, we initiate the study of the phase boundaries in the presence of interactions by locating the point where the many-body gap becomes minimum. The many-body spectral gap, denoted by $\Delta_g$, is defined as $\Delta_{g}=E_1-E_0$, where $E_0$ and $E_1$ are the ground and first excited state in the many-body spectrum of the system calculated under PBC~\cite{mikhail2022quasiparticle,Zhou_2023_PRB}. 
The transition between different phases is marked by the closing and subsequent reopening of $\Delta_g$ as we vary the Hamiltonian parameters.

\begin{figure}[htbp]
    \centering
    \includegraphics[width=1\linewidth]{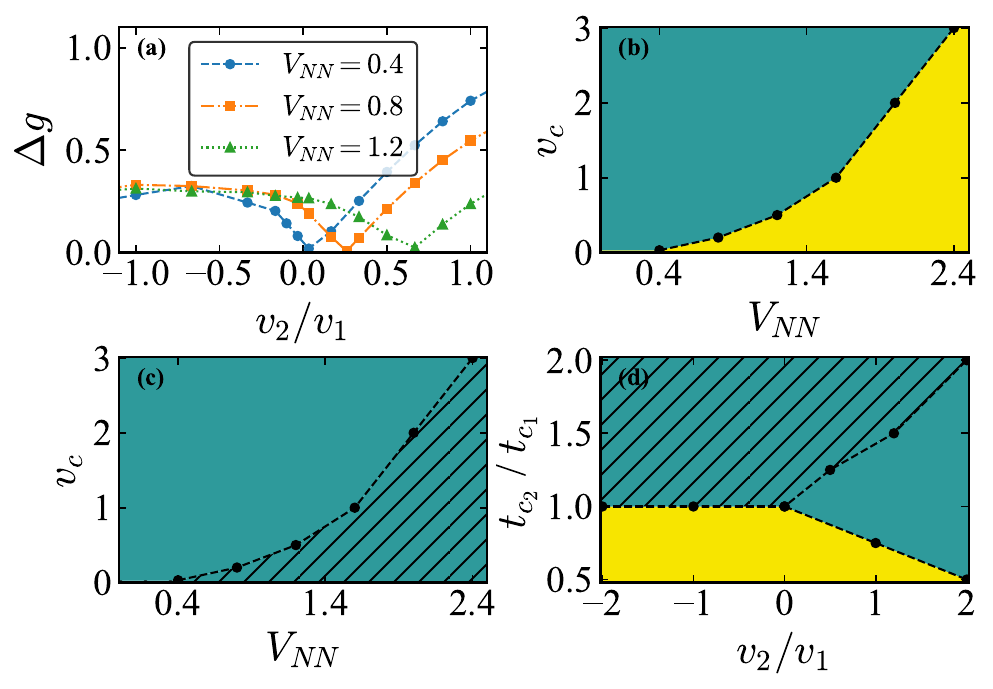}
    \caption{{\hl{The many-body spectral gap $\Delta_g = E_1 - E_0$ is obtained under PBC using DMRG, and its minimum signals the phase transition.}} Subfigure (a) shows the variation of $\Delta_g$ as a function of $v_2/v_1$ for $t_{c_2}/t_{c_1}=0.5$ and $V_{NN}=0.4,\, 0.8, \, 1.2$. A spectral gap minimum occurring at a critical value of $v_2/v_1$ (say $v_c$) signifies the location of the transition point, which shifts to higher values of $v_2/v_1$ as $V_{NN}$ increases. Panels (b,c) illustrate the shift in $v_c$ as a function of $V_{NN}$ for  $(t_{c_2}/t_{c_1}=0.5$)(in panel(b)) and ($t_{c_2}/t_{c_1} = 2$)(in panel(c)). In panel(d), we plot the critical boundaries, derived using the spectral gap analysis shown in panel(a), for $V_{NN}=2.0$ in the $(t_{c_2}/t_{c_1})-(v_2/v_1)$ plane. 
    The regions marked with solid colours yellow, green and green+hatches signify the different phases as identified by locating the spectral gap minima.
    Here, the lines connecting the markers serve as a guide to the eye. System size taken for this calculation is $L=40$.}
    \label{fig:gap_analysis_full}
\end{figure}

As a first step, we choose two different $t_{c_2}/t_{c_1}$ values, namely, $t_{c_2}/t_{c_1}=0.5$, lying deep inside the A phase in Figure~\ref{fig:Non-int_S_edge_by_ln2}, and $t_{c_2}/t_{c_1}=2.0$, that lies deep inside the C phase in Figure~\ref{fig:Non-int_S_edge_by_ln2}, when $V_{NN}=0$. We then scan across the $v_2/v_1$ axis in the presence of non-zero $V_{NN}$ to understand how the A-B or C-B non-interacting phase boundary occurring at $v_2/v_1=0$ transforms as we include interactions. We plot this study in Figure~\ref{fig:gap_analysis_full}. We identify the point where $\Delta_g$ is a minimum and define it as $v_c$. Figure~\ref{fig:gap_analysis_full}(a) shows that at $t_{c_2}/t_{c_1}=0.5$, the critical $v_2/v_1=v_c$ monotonically shifts to higher values as $V_{NN}$ increases from values, $V_{NN}=0.4, \, 0.8, \, 1.2$. A consolidated plot illustrating this trend of monotonic increase of $v_c$ with increasing $V_{NN}$ is shown in Figure~\ref{fig:gap_analysis_full}(b) for $t_{c_2}/t_{c_1}=0.5$. Thus, Figure~\ref{fig:gap_analysis_full}(b) highlights the transformation of the A-B (non-interacting) phase boundary, but now for $V_{NN}\ne 0$. We repeat a similar analysis for $t_{c_2}/t_{c_1}=2.0$ and identify a similar trend of increasing $v_c$ as $V_{NN}$ is included. Thus, the non-interacting phase boundary separating A and C phases from the B phase, occurring at $v_c=0$, now shifts to higher $v_c$ for $V_{NN}\ne0$. An alternative depiction of this result can be interpreted if we follow the $t_{c2}/t_{c1}=2.0$ or $t_{c2}/t_{c1}=0.5$ and $v_2/v1>0$ cuts in Figure~\ref{fig:gap_analysis_full}(d), where we plot the phase diagram in the $t_{c2}/t_{c1}$-$v_2/v_1$ plane for $V_{NN}=2.0$. Clearly, from Figure~\ref{fig:EE_for_U}, we had already identified that the A-C phase boundary remains unperturbed even in the presence of interactions. In Figure~\ref{fig:gap_analysis_full}(d), we assimilate all these results to derive the modified phase diagram when $V_{NN}\ne0$, where the non-trivial effect of interactions manifests as a monotonic shift of the A-B or C-B phase boundary (in Figure~\ref{fig:Non-int_S_edge_by_ln2}) to higher values of $v_2/v_1$ while the A-C phase boundary remains robust. For all the above calculations discussed in this section, we used a system size of $L=40$ with PBC. Within the system sizes checked, the drift of the gap-minimum position
was smaller than our numerical resolution. The gap-minimum loci define three interacting phase boundaries meeting at $v_2/v_1=0$ and $t_{c_2}/t_{c_1}=1$ within the resolution of the present calculations. The entanglement-spectrum analysis below assigns the corresponding phase character to the regions separated by these boundaries.

\subsubsection{Entanglement Spectrum}
\label{sec:entanglement_spectrum}

\begin{figure}[htbp]
    \centering
    \includegraphics[width=1\linewidth]{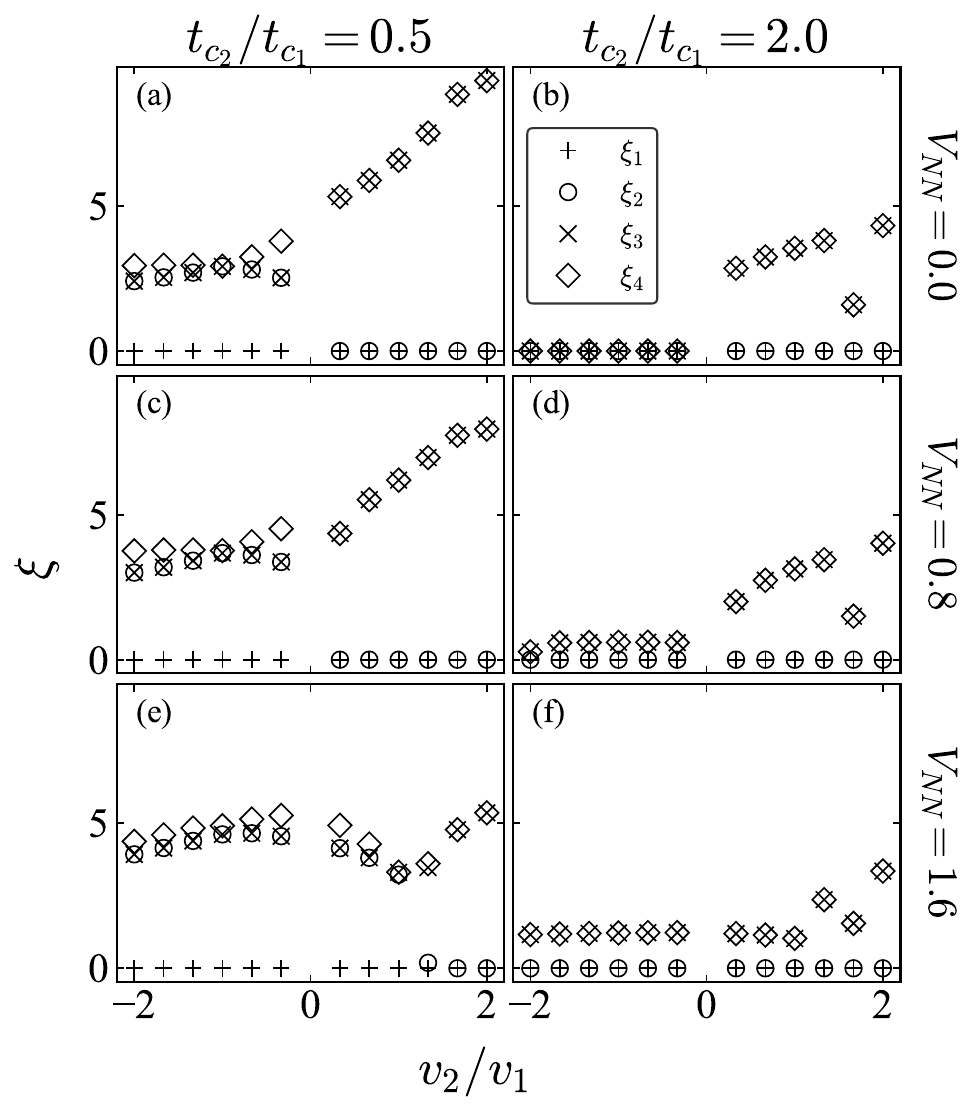}
    \caption{This figure illustrates scatter plots depicting the lowest four entanglement eigenvalues (\(\xi_{1,2,3,4}\)) as a function of \(v_2/v_1\). In the left (right) column we fix $t_{c_2}/t_{c_1}=0.5\,(2.0)$. In panels (a), (b) $V_{NN}=0.0$; (c), (d) $V_{NN}=0.8$; (e), (f) $V_{NN}=1.6$. The degeneracy of the lowest entanglement eigenvalue helps us to identify the topological and trivial phases. In the presence of interactions, the degeneracy of the lowest entanglement eigenvalue is identified as 1 (trivial) and 2 (topological), unlike the non-interacting case, where we also obtained 4-fold degeneracy as shown in panel (b). We have chosen a system size $L=512$ for this computation. (See main text for more details.)}
    \label{fig:E_spectrum}
\end{figure}

With the phase boundaries located from the many-body gap, we now use the low-lying entanglement spectrum to characterize the corresponding regions. Edge entanglement already provides this information for the $v_2/v_1<0$ cuts; the entanglement spectrum extends the analysis across the full range of $v_2/v_1$ considered here. The edge entanglement entropy calculation, for $v_2/v_1<0$, did help us in characterizing the topological and trivial phases as the $S_{edge}/\ln{2}$ value is directly correlated with the number of zero-energy edge states per edge present in the system. 
However, the same approach could not be adopted for $v_2/v_1>0$ in general. We thus resort to the entanglement spectrum analysis.

In Figure~\ref{fig:E_spectrum}, we present six scatter plots illustrating the low-lying entanglement spectrum, denoted as \(\{\xi_n\}\), plotted against the ratio $v_2/v_1$ for two different values  $t_{c_2}/t_{c_1}$. The first column corresponds to $(t_{c_2}/t_{c_1} = 0.5$, while the second column corresponds to $t_{c_2}/t_{c_1} = 2$. The interaction parameter $V_{NN}$ varies across the rows. The first row represents $V_{NN} = 0$, the second row $V_{NN} = 0.8$, the third row $V_{NN} = 1.6$.

The crucial idea in an entanglement spectrum analysis involves identifying the degeneracy of the lowest states in the spectrum. As shown in Figure~\ref{fig:E_spectrum}, For the case of $t_{c_2}/t_{c_1} = 0.5$, the degeneracy is one for all $v_2/v_1<0$, and for all $V_{NN}$. A non-degenerate low-lying state in the entanglement spectrum correlates with a trivial state~\cite{Li_2008_PRL_Entanglement} and hence the parameter set $t_{c2}/t_{c1}=0.5$ and $v_2/v_1<0$ therefore resides in a topologically trivial phase consistent with our previous analyses. This should prevail for all $V_{NN}$ and for all $t_{c2}/t_{c1}<1.0$, and $v_2/v_1<0$. 

In contrast, for $t_{c_2}/t_{c_1} = 2$, a degenerate entanglement spectrum ground state is found for all $v_2/v_1$, although the degeneracy value varies depending on the value of $v_2/v_1$ or $V_{NN}$. This is shown in Figure~\ref{fig:E_spectrum} (b, d, f). A degenerate ground state in the entanglement spectrum indicates that the system exhibits a topologically non-trivial state in the respective parameter regimes~\cite{Li_2008_PRL_Entanglement, Fidkowski_2010_PRL_Entanglement,Turner_2010_PRB_Entanglement,Chandran_2011_PRB_Bulk}. For $V_{NN} = 0$, i.e., the non-interacting case, this observation is consistent with the count of zero-mode end states, which is directly related to the winding number.

As interaction increases in the case of $t_{c_2}/t_{c_1} = 0.5$ Figure~\ref{fig:E_spectrum}(a,c,e), the initial value of $v_2/v_1$ at which degenerate states begin to emerge shifts to a higher value of $v_2/v_1$, thus opening up a topologically non-trivial state beyond $v_2/v_1>0$, unlike the non-interacting case. This observation aligns with the findings from the bulk gap analysis, Figure~\ref{fig:gap_analysis_full}. For the case of  $t_{c_2}/t_{c_1} = 2$, Figure~\ref{fig:E_spectrum}(b,d,f), the degeneracy of the lowest state changes from being 4-fold to 2-fold, even for $v_2/v_1<0$. However, for the $v_2/v_1>0$ part, the degeneracy of the lowest state remains the same as in the non-interacting case, i.e., 2-fold, even in the presence of interactions. Here, $t_{c_2}/t_{c_1}$ = 0.5 and the 2.0 are considered as representatives of parameters for $t_{c_2}/t_{c_1}<1.0$ and $t_{c_2}/t_{c_1}>1.0$, respectively.
This implies that when $V_{NN}\ne0$, the lowest eigenvalue in the entanglement spectra becomes 2-fold degenerate in the parameter space of phase $C$ of the non-interacting Hamiltonian. Note that this phase was 4-fold degenerate when $V_{NN}=0$. However, for the parameter space of phases A and B (in the non-interacting phase diagram), inclusion of interactions does not alter the degeneracy of the lowest state of the entanglement spectrum, i.e, the degeneracy remains the same, 1 and 2, respectively. 

\begin{figure}[htbp]
    \centering
    \includegraphics[width=1\linewidth]{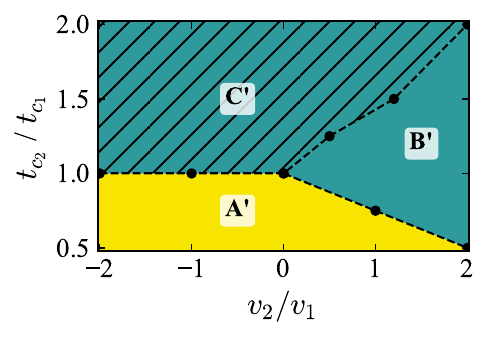}
    \caption{This figure represents the evolved phase diagram as we go beyond the non-interacting regime. Here, we choose $V_{NN}=2.0$. Phase A${}^\prime$ shows no degeneracy in the lowest eigenvalues of the entanglement spectrum. However, both the B${}^\prime$ and C${}^\prime$ phases show a two-fold degenerate lowest level of the entanglement spectrum. The phase boundaries are obtained using the spectral gap analysis as discussed in Figure~\ref{fig:gap_analysis_full}.}
    \label{fig:ES_evovled}
\end{figure}

Combining the gap-minimum locations with the entanglement-spectrum data gives the interacting phase diagram shown in Fig.~\ref{fig:ES_evovled} for $V_{NN}=2$. We denote the three connected gapped regions by $A'$, $B'$, and $C'$, according to the free phases to which they are adiabatically connected. The lowest entanglement level is nondegenerate in $A'$ and twofold degenerate in both $B'$ and $C'$. The many-body $\mathbb Z_4$ chiral index derived below distinguishes $B'$ and $C'$ despite their common twofold lowest entanglement level.

The interacting phase diagram now raises two complementary questions.
First, what are the many-body SPT labels of the three connected gapped
regions $A'$, $B'$, and $C'$? Second, what determines the different
responses of the $A'$-$C'$ and $B'$-$C'$ critical boundaries?
We address the SPT classification first and turn subsequently to the
critical theories of the two boundaries.
\subsubsection{Many-body chiral index of the interacting phases}
\label{sec:manybody_spt_index}

For the interacting half-filled problem the protecting symmetry is
Eq.~\eqref{eq:Gf_micro},
\begin{equation}
G_f=U(1)_Q\times\mathbb Z_2^{\mathcal S},
\,
\mathcal S i\mathcal S^{-1}=-i,
\,
\mathcal S Q\mathcal S^{-1}=-Q.
\end{equation}
One-dimensional charge-conserving fermionic SPT phases with this many-body chiral symmetry are classified by $\mathbb Z_4$
\cite{LeeShiozakiHsieh2024,BourneOgata2021}. We denote the index by
\begin{equation}
\Omega_{\mathcal S}
=
\exp\left(\frac{i\pi}{2}I_{\mathcal S}\right),
\qquad
I_{\mathcal S}\in\mathbb Z_4 .
\label{eq:IS_definition}
\end{equation}

The four classes can be seen directly from their edge symmetry
representations. Suppose a boundary carries $q$ elementary complex edge
modes $d_m$, $m=1,\ldots,q$, satisfying
\begin{equation}
\mathcal S_e d_m\mathcal S_e^{-1}=d_m^\dagger .
\end{equation}
Let
\begin{equation}
P_e=(-1)^{\sum_m d_m^\dagger d_m}
\end{equation}
be the edge fermion parity. A direct Clifford-algebra construction,
given in Appendix~\ref{app:manybody_spt}, yields
\begin{equation}
P_e\mathcal S_eP_e^{-1}=(-1)^q\mathcal S_e,
\qquad
\mathcal S_e^2=(-1)^{q(q-1)/2}.
\label{eq:edge_Z4_algebra_main}
\end{equation}
Consequently,
\begin{equation}
\begin{array}{c|c|c|c}
q\pmod 4
&
P_e\mathcal S_eP_e^{-1}
&
\mathcal S_e^2
&
I_{\mathcal S}=-q\pmod 4
\\ \hline
0 & +\mathcal S_e & +1 & 0\\
1 & -\mathcal S_e & +1 & 3\\
2 & +\mathcal S_e & -1 & 2\\
3 & -\mathcal S_e & -1 & 1
\end{array}
\label{eq:edge_Z4_table}
\end{equation}
The $q=1$ and $q=2$ phases are therefore inequivalent. In the first,
the localized chiral operation is fermion-parity odd; in the second it is parity even but squares projectively to $-1$. Both classes may exhibit a twofold lowest entanglement level, while their different edge symmetry algebras provide the $\mathbb Z_4$ distinction.

In the block-diagonal SSH representative of Appendix
~\ref{app:decoupled_SSH}, phase $A$ contains no topological SSH block,
phase $B$ contains one, and phase $C$ contains two. With the orientation
used for the winding numbers in this paper, $q=-\nu$. The free
representatives therefore carry
\begin{equation}
I_{\mathcal S}(A)=0,\qquad
I_{\mathcal S}(B)=-1,\qquad
I_{\mathcal S}(C)=-2
\pmod 4.
\label{eq:free_Z4_labels}
\end{equation}

The primed phases $A'$, $B'$, and $C'$ are defined as the connected
symmetry-preserving gapped components containing the corresponding free
phases $A$, $B$, and $C$. For any point in one of these components,
choose a continuous path
\begin{equation}
H_X(s),\qquad 0\leq s\leq1,
\end{equation}
that remains inside the same gapped component, with $H_X(0)$ chosen in
the free representative and $H_X(1)$ at the interacting point of
interest. Along this path the many-body gap remains finite and the
symmetry $G_f$ is preserved. Symmetry-preserving quasi-adiabatic
continuation therefore preserves the projective edge representation
\cite{pollmann2012symmetry,turner2011topological,HastingsWen2005}.

The numerical phase diagram provides representative points well inside
each of the three components. On the
$t_{f_2}/t_{f_1}=1$ cut, convenient choices are
\begin{equation}
p_A=(0.5,-1),\qquad
p_C=(2,-1),\qquad
p_B=(2,+3),
\end{equation}
where the entries denote
$(t_{c_2}/t_{c_1},v_2/v_1)$. Figures~\ref{fig:EE_for_U} and
\ref{fig:gap_analysis_full} place these points inside the corresponding
connected gapped regions throughout the interaction range used in the
phase diagram.
Hence
\begin{equation}
\boxed{
I_{\mathcal S}(A')=0,\,
I_{\mathcal S}(B')=-1,\,
I_{\mathcal S}(C')=-2
\pmod 4 .
}
\label{eq:interacting_Z4_labels}
\end{equation}

Across the $A'$-$C'$ boundary the adjacent gapped phases differ by
\begin{equation}
\Delta I_{\mathcal S}=2\pmod 4,
\end{equation}
while the interacting critical line has central charge $c=1$. These
quantities characterize different aspects of the problem:
$I_{\mathcal S}$ specifies the projective edge structure of the gapped phases, whereas $c$ counts the gapless bulk degrees of freedom at the critical point. Together they characterize an interacting transition between two distinct $\mathbb Z_4$ SPT sectors with a single critical collective degree of freedom. The distinction between the SPT index of the adjacent gapped phases and the conformal data of the intervening critical theory is worth noting. The former does not by itself fix the universality class of the transition. This viewpoint is consistent with analyses of interacting one-dimensional SPT transitions, where the projective edge structure of the gapped phases and the gapless critical degrees of freedom provide complementary information \cite{VerresenMoessnerPollmann2017}.

\paragraph*{Microscopic comparison of the two topological routes.}

The ladder contains two microscopic ingredients associated with its
topological edge structure: SSH-like dimerization and directional
$c$-$f$ hybridization. Their roles can be compared through two
representative parameter cuts.

\begin{figure}[htbp]
    \centering
    \includegraphics[width=1\linewidth]{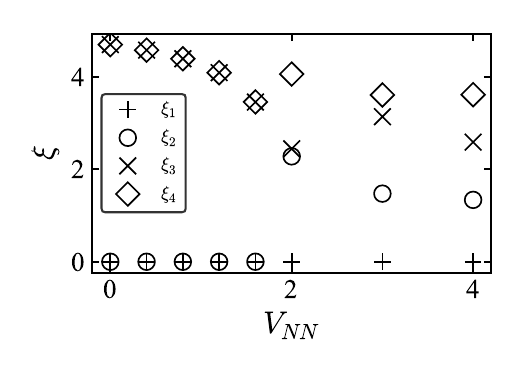}
    \caption{Lowest four entanglement eigenvalues along the symmetric
    $p$-wave-periodic-Anderson-inspired cut
    $t_{c_1}=t_{c_2}$, $t_{f_1}=t_{f_2}$, and $v_1=v_2$, as a function
    of $V_{NN}$. The interaction progressively lifts the lowest
    entanglement-spectrum degeneracy along this cut. The system size is
    $L=512$.}
    \label{fig:p_PAM_E_spectrum}
\end{figure}

The symmetric cut
\begin{equation}
t_{c_2}=t_{c_1},\qquad
t_{f_2}=t_{f_1},\qquad
v_2=v_1
\end{equation}
is the $p$-wave-periodic-Anderson-inspired limit discussed in
Sec.~\ref{sec:model}. Figure~\ref{fig:p_PAM_E_spectrum} shows that the
lowest entanglement-spectrum degeneracy is progressively lifted by the
density interaction along this cut.

A complementary dimerized cut is obtained by taking
$t_{c_2}/t_{c_1}>1$ at $t_{f_2}/t_{f_1}=1$. For example,
Fig.~\ref{fig:E_spectrum}(b,d,f) shows that the point
$v_2/v_1=1$ retains a twofold lowest entanglement level across the
sampled interaction strengths. As the $B'$-$C'$ boundary moves with
$V_{NN}$, this point passes between the two distinct interacting
topological sectors identified by their $\mathbb Z_4$ indices. The
comparison therefore shows that SSH-like dimerization provides a robust
route to interacting topological edge structure for the density
interaction studied here.

Having characterized the interacting gapped phases, we next turn to the
critical degrees of freedom on their boundaries.
\subsubsection{Central Charge}
\label{sec:results_int_c_charge}

To characterize the gapless degrees of freedom on the critical lines,
we extract the central charge from the scaling of the bipartite
entanglement entropy. For a periodic $(1+1)$-dimensional conformal field
theory of spatial length $L$, the ground-state entropy of an interval of
length $l$ obeys the Calabrese-Cardy form
\begin{equation}
S(l)
=
\frac{c}{3}
\log\left[
\frac{L}{\pi}
\sin\left(\frac{\pi l}{L}\right)
\right]
+s_0 ,
\label{cc}
\end{equation}
where $s_0$ is a nonuniversal constant. For OBC the coefficient of the
logarithm is $c/6$~\cite{Calabrese:2004eu}.

For the free $A$-$C$ transition, correlation-matrix calculations on
systems of size $L\sim2000$ give $c\simeq2$. This agrees with the two
gap-closing Dirac valleys at $\pm k_0$ derived in
Sec.~\ref{sec:free_critical_content}. The free gapped phase $C$ is
independently characterized by winding number $\nu=-2$ and
$S_{\rm edge}=2\ln2$.

With interaction, the DMRG fits along the $A'$-$C'$ critical line
evolve toward $c=1$, as shown in Fig.~\ref{fig:cc_EE}. The low-energy
projection in Appendix~\ref{app:AC_projection} shows that the density
interaction couples the two valley sectors. In the analytically
transparent weak-coupling limit discussed there, this coupling produces
one massive and one gapless collective combination, providing a
continuum realization of the observed $c=1$ critical theory. At weak
interaction the finite-size crossover is slow, consistent with a
correlation length comparable to or larger than the available system
sizes.

The free $B$-$C$ transition contains one light Dirac species at
$k=\pi$. The self-energy analysis in
Sec.~\ref{sec:BC_shift_main} shifts the massless condition to
$v_2=v_{2,c}(V_{NN})$ while retaining one light Dirac sector in the
weak-coupling theory. Its critical content is therefore that of a
$c=1$ Dirac theory at the shifted point.

\begin{figure}[htbp]
    \centering
    \includegraphics[width=1\linewidth]{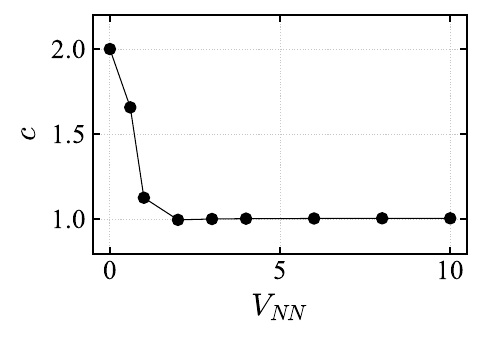}
    \caption{Central charge $c$ as a function of interaction calculated on the transition line $t_{c_2}/t_{c_1}=1$ at  $v_2/v_1=-2$ with system size $L=120$. The solid line connecting the points is intended to serve as a guide for the eye.}
    \label{fig:cc_EE}
\end{figure}

\subsection{Free critical content of the two transition lines}
\label{sec:free_critical_content}
In this section we discuss the effective low energy physics in the model using a continuum approach. 

\paragraph*{$A$-$C$ boundary.}
Set
\begin{equation}
t_{f_1}=t_{f_2}\equiv t_f,\qquad
\bar t_c=\frac{t_{c_1}+t_{c_2}}{2},\qquad
\delta t_c=\frac{t_{c_1}-t_{c_2}}{2}.
\end{equation}
Using Eq.~\eqref{eq:h(k)}, the determinant of the chiral block is
\begin{align}
\det h(k)
&=
-4e^{-ik}
\Bigg[
\bar t_c t_f\cos^2\frac{k}{2}
+
v_1v_2\sin^2\frac{k}{2}
\nonumber\\
&\hspace{3.2cm}
+i t_f\delta t_c
\sin\frac{k}{2}\cos\frac{k}{2}
\Bigg].
\label{eq:AC_determinant}
\end{align}
For a generic closing at $k_0\neq0,\pi$, the real and imaginary parts
must vanish independently:
\begin{equation}
\delta t_c=0,
\qquad
\tan^2\frac{k_0}{2}
=
-\frac{\bar t_c t_f}{v_1v_2}.
\label{eq:AC_closing_condition}
\end{equation}
A real solution $k_0\in(0,\pi)$ exists provided $\bar t_c t_f\,v_1v_2<0$. This condition holds on all the interacting cuts studied here: with $t_{c_1}=1$, $t_f=0.3$, $v_1=-0.3$ and $v_2/v_1<0$, one has $\bar t_c t_f>0$ and $v_1v_2<0$. The free $A$-$C$ boundary is $t_{c_1}=t_{c_2}$, and it contains two gap-closing valleys $k_\lambda=\lambda k_0$, $\lambda=\pm$. The free critical theory consequently has two Dirac flavours and $c=2$.

Projecting onto the two light states at each valley gives the two-flavour Dirac theory
\begin{align}
H_{AC}^{(0)}
&=
\sum_{\lambda=\pm}\int dx\,
\psi_\lambda^\dagger
\left(
-i\lambda v_*\sigma_z\partial_x
+
m_*\sigma_x
\right)
\psi_\lambda,
\nonumber\\
\psi_\lambda
&=
\begin{pmatrix}
R_\lambda & L_\lambda
\end{pmatrix}^{T},
\qquad
m_*\propto \delta t_c .
\label{eq:HAC_Dirac_main}
\end{align}
Thus the microscopic dimerization $\delta t_c$ is the common Dirac
mass of the two valleys. The explicit construction of the light states
and the corresponding projected coefficients are given in
Appendix~\ref{app:AC_projection}.

\paragraph*{$B$-$C$ boundary.}
The $k=\pi$ condition in Eq.~\eqref{eq:kpi_boundary} reduces on
$t_{f_1}=t_{f_2}$ to $v_2=0$ for $v_1\neq0$. The gap closes at the
single momentum $k=\pi$, and expansion about this point gives the
one-flavour Dirac Hamiltonian,
\begin{equation}
H_{BC}^{(0)}=\int dx\,\psi^\dagger\left(-iv_{BC}\sigma_z\partial_x+m_{BC}\sigma_x\right)\psi,
\qquad
m_{BC}\propto v_2,
\label{eq:HBC_Dirac_main}
\end{equation}
so the free $B$-$C$ line has $c=1$.

In the following, we argue why one of the phase boundaries remains pinned while the other shifts in the presence of density interactions.

\subsubsection{Pinning of the
\texorpdfstring{$A$-$C$}{A-C} boundary}
\label{sec:pin_shift_mechanism}

The pinned location of the $A'$-$C'$ transition follows from an exact reflection property of the microscopic Hamiltonian. On an infinite chain, or under PBC after a relabelling of the unit cells,
define
\begin{align}
\mathcal R c_{a,j}\mathcal R^{-1}&=c_{a,-j},
&
\mathcal R c_{b,j}\mathcal R^{-1}&=c_{b,-j-1},
\nonumber\\
\mathcal R f_{a,j}\mathcal R^{-1}&=-f_{a,-j},
&
\mathcal R f_{b,j}\mathcal R^{-1}&=-f_{b,-j-1}.
\label{eq:AC_reflection_action}
\end{align}
A direct transformation of Eq.~\eqref{eq:H0} gives
\begin{align}
&\mathcal R
H(t_{c_1},t_{c_2},t_{f_1},t_{f_2},v_1,v_2,V_{NN})
\mathcal R^{-1}
\nonumber\\
&\qquad=
H(t_{c_2},t_{c_1},t_{f_2},t_{f_1},v_1,v_2,V_{NN}).
\label{eq:reflection_parameter_exchange}
\end{align}
The density interaction is invariant under this transformation. On the cut $t_{f_1}=t_{f_2}$, reflection therefore sends
\begin{equation}
\delta t_c=\frac{t_{c_1}-t_{c_2}}{2}
\longrightarrow
-\delta t_c .
\label{eq:reflection_delta_tc}
\end{equation}
Hence the phase diagram on this cut is symmetric about $\delta t_c=0$. The numerical gap and entanglement results show a single, unsplit $A'$-$C'$ transition. Reflection then fixes this transition at its invariant point,
\begin{equation}
\delta t_c=0,
\qquad\text{i.e.}\qquad
t_{c_1}=t_{c_2}.
\end{equation}

The change in the critical central charge is a separate property of this transition. The free line contains two Dirac valleys and has $c=2$, whereas the interacting central-charge calculation in Fig.~\ref{fig:cc_EE} approaches $c=1$. The low-energy projection in Appendix~\ref{app:AC_projection} shows that the density interaction couples the two valley sectors and provides a weak-coupling continuum interpretation of this reduction.

\begin{greenblock}
    \subsubsection{Shifting of the \texorpdfstring{$B$-$C$}{B-C} boundary}
\label{sec:BC_shift_main}

At the free $B$-$C$ boundary, set $v_2=0$ and expand around $k=\pi$.
For $t_{f_1}=t_{f_2}$, define
\begin{equation}
m_c=t_{c_1}-t_{c_2},
\qquad
M=\sqrt{m_c^2+4v_1^2}.
\end{equation}
The light orbital combinations may be chosen as
\begin{equation}
d_j=\frac{2v_1c_{a,j}-m_cf_{a,j}}{M},
\qquad
e_j=f_{b,j}.
\label{eq:BC_light_fields}
\end{equation}
Their slowly varying components form a single Dirac field, and the
quadratic perturbation $v_2$ couples directly to its mass. With
$G^{-1}=G_0^{-1}-\Sigma$, the off-diagonal element of the inverse Green
function projected to the light subspace is
\begin{align}
[K_{LL}(0,\pi)]_{12}
&=
\frac{4v_1}{M}v_2
-\frac{2v_1}{M}\Sigma_{14}^{(2)}(0,\pi)
\nonumber\\
&\quad
+\frac{m_c}{M}\Sigma_{24}^{(2)}(0,\pi)
+O(V_{NN}^3).
\label{eq:BC_projected_kernel}
\end{align}
Here the orbital indices refer to the basis $(c_a,f_a,c_b,f_b)$ used in Eq.~\eqref{eq:chiral_basis_hamiltonian}. Thus $\Sigma_{14}$ is the interaction-generated $c_a^\dagger f_b$ hybridization channel, while $\Sigma_{24}$ is the $f_a^\dagger f_b$ channel. Both contribute to the low-energy Dirac mass because the light field $d$ in Eq.~\eqref{eq:BC_light_fields} contains both $c_a$ and $f_a$ components. The origin of the projection coefficients and the explicit perturbative calculation of these self-energies are given in Appendix~\ref{app:BC_self_energy}.

At first order, the Hartree and Fock contractions populate diagonal and
same-sublattice light-heavy channels. The Dirac-mass channel appears
at second order through virtual heavy-mode processes, giving the
interacting critical condition
\begin{equation}
\boxed{
2v_{2,c}(V_{NN})
=
\Sigma_{14}^{(2)}(0,\pi)
-
\frac{m_c}{2v_1}
\Sigma_{24}^{(2)}(0,\pi)
+
O(V_{NN}^3).
}
\label{eq:v2_shift_main}
\end{equation}
Both self-energies are $O(V_{NN}^2)$, so the leading displacement is
quadratic in $V_{NN}$. Appendix~\ref{app:BC_self_energy} derives the
light-subspace projection and the corresponding lattice self-energy
integrals explicitly. The shifted massless point retains the single
light Dirac sector of the free $B$-$C$ theory.
\end{greenblock}

\subsection{Interaction-class viewpoint}
\label{sec:interaction_class_viewpoint}
The projected operator content provides a useful organization of the two boundary responses. At the $A$-$C$ line, reflection fixes the observed unsplit transition at $\delta t_c=0$. The density interaction also couples the two valley sectors, while the numerical central-charge analysis shows that the interacting critical theory has $c=1$. Appendix~\ref{app:AC_projection} gives a weak-coupling field-theory realization of this reduction.

At the $B$-$C$ line, the critical theory contains a single Dirac species. The density interaction studied here generates its leading mass shift at second order, as in Eq.~\eqref{eq:v2_shift_main}. Interactions coupling directly to the same Dirac-mass channel, such as correlated hybridization terms, can produce a first-order displacement. Thus the symmetry and scaling content of the projected operators determines how a given microscopic interaction reshapes the critical boundary. It is useful to state the scope of this conclusion explicitly. On the cut $t_{f1}=t_{f2}$, any additional interaction that preserves both the protecting symmetry and the reflection action of \eqref{eq:reflection_parameter_exchange} gives a phase diagram symmetric under $\delta t_c\to-\delta t_c$. Therefore, provided the $A'-C'$ boundary remains a single unsplit direct transition, reflection pins it to $\delta t_c=0$. Reflection symmetry alone does not exclude a pair of transitions at $\pm\delta t_c^{\star}$ with an intervening phase, and an interaction that breaks this reflection can shift the boundary. The same reflection does not pin the $B'-C'$ boundary because its tuning mass is proportional to $v_2$, which is invariant under \eqref{eq:reflection_parameter_exchange}, rather than odd under the reflection. Thus \eqref{eq:reflection_parameter_exchange} constrains the $t_{c1}-t_{c2}$ direction but does not require $v_{2,c}$ to vanish. More generally, the order at which a boundary moves is interaction dependent. For \eqref{eqn:H_int1} the $B'-C'$ mass is generated only at $O(V_{\rm NN}^{2})$, whereas a symmetry-allowed interaction with a nonzero direct projection onto the same Dirac-mass channel may shift it already at first order.
\section{Conclusions}
\label{sec:conclusions}

We studied the interacting topological phase diagram of a spinless two-leg ladder with directional, $p$-wave-like inter-orbital hybridization. { The motivation behind designing such a model is two fold. Primarily, it provides a minimal and controlled setting to probe the effects of interactions on topological phase boundaries with different low energy critical content. It also bears close resemblance with topological Kondo insulator Hamiltonians. 

We showed that, in the non-interacting limit, the model supports three insulating regions with winding numbers $\nu=0,-1,-2$, namely the $A$, $B$, and $C$ phases, respectively. Our analytical, continuum calculations, show that the $A$-$C$ boundary contains two Dirac valleys and has $c=2$ in the free theory. 
%On the cut $t_{f_1}=t_{f_2}$, reflection exchanges $t_{c_1}$ and $t_{c_2}$ and sends $\delta t_c\rightarrow-\delta t_c$. 
On the other hand, the $B$-$C$ boundary has a different low-energy structure, with a single Dirac mode at $k=\pi$. This provides a unique setting where the effects of interactions on the topological phase boundaries, with qualitatively different low energy critical theories, can be investigated within the same Hamiltonian. Our numerical many-body calculations, using DMRG and entanglement diagnostics 
%based on edge entanglement, entanglement spectra, many-body gap minima, and central-charge fits 
show that the same local density interaction affects the two critical boundaries in qualitatively different ways. While the $A$-$C$ boundary remains pinned, the $B$-$C$ boundary shifts in a particular as interactions are included. Furthermore, the pinning of the $A$-$C$ boundary does not imply that it is unaffected by interactions. On the contrary, its low energy critical structure is substantially modified by interactions. We showed how the fate of these distinct topological boundaries is determined by the symmetry and valley structure of the critical theory. This is one of the main results of this work.}

The interacting phases $B'$ and $C'$ therefore represent distinct SPT classes, although both exhibit a twofold lowest entanglement level in the parameter regime studied here. The reduction of the lowest entanglement degeneracy in the region connected to the free $\nu=-2$ phase is consistent with the restructuring of the edge multiplet by interactions. 
Along the $A$-$C$ boundary, that has $c=2$ in the free theory, on the cut $t_{f_1}=t_{f_2}$, reflection exchanges $t_{c_1}$ and $t_{c_2}$ and sends $\delta t_c\rightarrow-\delta t_c$. Together with the numerically observed single transition, this symmetry fixes the interacting $A'$-$C'$ boundary at $t_{c_1}=t_{c_2}$. Independently, the central-charge fits show that the interacting critical theory approaches $c=1$. The continuum projection, and the weak-coupling analysis show that the density interaction couples the two valley sectors, 
concomitant with the loss of one critical collective degree of freedom.

%The continuum projection shows that the density interaction couples the two valley sectors, and the weak-coupling analysis 
%provides a microscopic realization consistent with the loss of one critical collective degree of freedom.

Along the $B$-$C$ boundary the leading interaction-induced renormalization of its Dirac mass arises at second order through virtual heavy modes. The resulting self-energy shifts the critical point according to Eq.~\eqref{eq:v2_shift_main}, while the transition retains its one-Dirac, $c=1$ critical content.
Together, these results show that the response of a topological phase boundary is governed by the symmetry and operator content of its projected critical theory. In the present model, reflection fixes the location of the $A'$-$C'$ transition, intervalley interactions reorganize its critical degrees of freedom, and heavy-mode self-energies shift the $B'$-$C'$ transition. 

{We also associate a SPT label for each of the interacting topological phases.}
At half filling, charge conservation together with the antiunitary
many-body chiral symmetry $\mathcal S$ gives the interacting
$\mathbb Z_4$ classification, namely, 
%\begin{equation}
$I_{\mathcal S}(A')=0$,$I_{\mathcal S}(B')=-1$, $I_{\mathcal S}(C')=-2
\pmod 4$.
Extensions to other short-range interactions, spinful versions of the ladder, disorder, and higher-dimensional settings provide natural directions for further study.

\section{Methods}
\label{sec:methods}

We study the ladder model shown in Fig.~\ref{fig:model} under interaction defined in Eq.~\eqref{eqn:H_int1}, at half filling. For non-interacting $(V_{NN}=0)$ case, we calculate the winding number ($\nu$) from the phase of $\det h(k)$ and obtain the Edge entanglement entropy as defined in Eq.\eqref{edge_def}.

For interacting case $(V_{NN}\neq 0)$, we map the ladder onto a one-dimensional ordering of sites and calculate its ground state using density-matrix renormalization group (DMRG)~\cite{white1992density,Schollwock2011}. We have considered Bond dimensions up to $\chi=1500$ and convergence check of the energy and entropy to $10^{-8}$. Unless stated otherwise, we set $t_{c_1}=1$, $v_1=-0.3$, and $t_{f_1}=t_{f_2}=0.3$. The system size $L$ counts microscopic orbitals, so $L=4L_{\rm uc}$; sizes used for individual results are given with the corresponding figures.

We have calculated Edge entanglement entropy for Interacting case as well. We estimate transition points from minima of the finite-size many-body gap $\Delta_g=E_1-E_0$ under PBC, and characterize the gapped (both topological and trivial) regions using the low-lying entanglement spectrum. Along $A$-$C$ critical line Fig.~\ref{fig:Non-int_S_edge_by_ln2}, we calculate the central charge by fitting the PBC entropy to Eq.~\eqref{cc} (Details: Appendix~\ref{app:central_charge}). The analytical gap-closing conditions, low-energy projection, and perturbative self-energy calculation are detailed in Appendices~\ref{app:exact_calc}, \ref{app:AC_projection}, and \ref{app:BC_self_energy}, respectively.

\bmsubsection*{Acknowledgments}
AH acknowledges support from ANRF, India (Project No. SRG/2022/000495), IIT(ISM) Dhanbad, India, and HPC (Aryabhatta) facility of IIT(ISM) Dhanbad. SS acknowledges support from the ANRF, India (Grant Nos. SRG/2022/000495 and MTR/2022/000638) and IIT(ISM) Dhanbad (Project No. FRS(175)/2022-2023/PHYSICS). We would also like to acknowledge Prof. Awadhesh Narayan (IISc) and Prof. Tapan Mishra (NISER) for insightful discussions during the initial stage of the work. NA would like to thank the Start-up Research Grant (SRG/2022/000972) for the computational facilities at BITS, and Sharanga HPC usage for tensor network computations. NA further acknowledges Sharanga HPC usage for tensor network computations, Kapil Ghadiali for computational support at TIFR and hospitality at the Institute for Basic Science, Daejeon, where a part of the work was done. RB acknowledges support from ANRF, India, for support through grant MTR/2022/000795 and BITS Pilani for support through grant CDRF: C2/24/282.

\bmsubsection*{Conflict of Interest}
The authors declare no conflict of interest.

\bmsubsection*{Data Availability Statement}
The data and code supporting the findings of this study are available
from the corresponding authors upon request.

% \clearpage
\setcounter{lastmainfigure}{\value{figure}}
\begin{appendix}
\setcounter{figure}{\value{lastmainfigure}}
\renewcommand{\thefigure}{\arabic{figure}}
\numberwithin{equation}{section}
%%%%%%%%%%%%%%%%%%%%%%%%%%Appendix A%%%%%%%%%%%%%%%%%%%%%%%%%%%%%%

\begin{greenblock}
    \section{Analytical solution of the non-interacting Hamiltonian}
\label{app:exact_calc}

\subsection{Spectrum and gap-closing conditions}

Write the off-diagonal block as
\begin{equation}
h(k)=
\begin{pmatrix}
p_k&s_k\\
q_k&r_k
\end{pmatrix},
\end{equation}
where
\begin{align}
p_k&=t_{c_1}+t_{c_2}e^{-ik},
&
q_k&=v_1(1-e^{-ik}),
\nonumber\\
r_k&=-(t_{f_1}+t_{f_2}e^{-ik}),
&
s_k&=v_2(e^{-ik}-1).
\label{eq:appendix_h_elements}
\end{align}
The squared Bloch Hamiltonian is
\begin{equation}
\mathcal H_k^2
=
\begin{pmatrix}
h(k)h^\dagger(k)&0\\
0&h^\dagger(k)h(k)
\end{pmatrix},
\end{equation}
so the positive energies are the two singular values of $h(k)$. Define
\begin{equation}
T_k
=
|p_k|^2+|q_k|^2+|r_k|^2+|s_k|^2,
\qquad
\mathcal D_k
=
|\det h(k)|^2 .
\end{equation}
The four bands are
\begin{align}
E_{\pm,1}(k)
&=
\pm
\sqrt{
\frac{T_k}{2}
-
\sqrt{\frac{T_k^2}{4}-\mathcal D_k}
},
\nonumber\\
E_{\pm,2}(k)
&=
\pm
\sqrt{
\frac{T_k}{2}
+
\sqrt{\frac{T_k^2}{4}-\mathcal D_k}
}.
\label{eq:free_singular_values}
\end{align}
The bulk gap closes exactly when
\begin{equation}
\mathcal D_k=|\det h(k)|^2=0.
\end{equation}
For real hopping and hybridization amplitudes,
\begin{align}
0
&=
\Big[
t_{c_2}t_{f_1}
+t_{c_1}t_{f_2}
+2v_1v_2
\nonumber\\
&\qquad+
\left(
t_{c_1}t_{f_1}
+t_{c_2}t_{f_2}
-2v_1v_2
\right)\cos k
\Big]^2
\nonumber\\
&\quad+
\left(
t_{c_1}t_{f_1}
-t_{c_2}t_{f_2}
\right)^2\sin^2 k .
\label{eq:Main_condition}
\end{align}
Since this is a sum of nonnegative terms, each term must vanish.

At $k=\pi$ one obtains
\begin{equation}
(t_{c_2}-t_{c_1})(t_{f_2}-t_{f_1})
=
4v_1v_2,
\label{eq:kpi_boundary}
\end{equation}
or
\begin{equation}
\frac{t_{c_2}}{t_{c_1}}
=
1+
\frac{
4v_1^2
}{
(t_{f_2}/t_{f_1}-1)t_{f_1}t_{c_1}
}
\frac{v_2}{v_1}.
\label{eq:oblique_boundary}
\end{equation}
The corresponding angle in Fig.~\ref{fig:PhaseDiagram_nu} is
\begin{equation}
\theta
=
\tan^{-1}
\left[
\frac{
4v_1^2
}{
(t_{f_2}/t_{f_1}-1)t_{f_1}t_{c_1}
}
\right].
\end{equation}

At $k=0$, the condition reduces to
\begin{equation}
(t_{c_1}+t_{c_2})(t_{f_1}+t_{f_2})=0.
\end{equation}
This high-symmetry branch is outside the generic parameter region used
for the interacting phase diagram.

For a generic closing with $\sin k\neq0$,
\begin{equation}
t_{c_1}t_{f_1}=t_{c_2}t_{f_2},
\label{eq:generic_horizontal_condition}
\end{equation}
together with
\begin{equation}
\cos k
=
-
\frac{
t_{c_2}t_{f_1}
+t_{c_1}t_{f_2}
+2v_1v_2
}{
t_{c_1}t_{f_1}
+t_{c_2}t_{f_2}
-2v_1v_2
},
\qquad
|\cos k|\leq1.
\label{eq:generic_closing_momentum}
\end{equation}
Equation~\eqref{eq:generic_horizontal_condition} gives the horizontal
line
\begin{equation}
\frac{t_{c_2}}{t_{c_1}}
=
\frac{t_{f_1}}{t_{f_2}}
\end{equation}
in the phase diagrams.

\subsection{Flattened Hamiltonian and winding number}

The explicit Bloch eigenvectors are not required to compute the winding.
For a gapped momentum define the spectrally flattened Hamiltonian
\begin{equation}
Q(k)
=
\operatorname{sgn}\mathcal H_k
=
\mathcal H_k\left(\mathcal H_k^2\right)^{-1/2}.
\end{equation}
Chiral symmetry implies
\begin{equation}
Q(k)
=
\begin{pmatrix}
0&q(k)\\
q^\dagger(k)&0
\end{pmatrix},
\qquad
q(k)
=
h(k)\left[h^\dagger(k)h(k)\right]^{-1/2}.
\label{eq:flattened_q}
\end{equation}
The matrix $q(k)$ is unitary whenever $\det h(k)\neq0$. The winding
number used in the phase diagram is
\begin{align}
\nu
&=
\frac{1}{2\pi i}
\int_{-\pi}^{\pi}dk\,
\operatorname{Tr}
\left[
q^\dagger(k)\partial_kq(k)
\right]
\nonumber\\
&=
\frac{1}{2\pi i}
\int_{-\pi}^{\pi}dk\,
\partial_k\log\det h(k).
\label{eq:winding_polar}
\end{align}
The positive Hermitian factor in the polar decomposition has a real
positive determinant and therefore does not contribute to the phase
winding. Hence every free topological transition occurs at a zero of $\det h(k)$, precisely as obtained above.
\end{greenblock}
\section{Special cases of the non-interacting Hamiltonian}
\label{app:decoupled_SSH}
In this section, we discuss a special limit, where the non-interacting model simply maps onto two independent SSH chains.
We rewrite the Hamiltonian Equation~\eqref{eq: pauli_decomposed_hamiltonian} for the case, $v_1 = -v_2$, such that, in the $\sigma \otimes \tau$ basis, now, it takes the following form:
\begin{align}
    \mathcal{H}^a_k &= 
    \big[(t_{cf1}+t_{cf2}\cos k)\,\sigma_x + t_{cf2}\sin k\,\sigma_y\big]\otimes \tau_0 \nonumber\\
    &\quad + \big[(t_{cf1}'+t_{cf2}'\cos k)\,\sigma_x + t_{cf2}'\sin k\,\sigma_y\big]\otimes \tau_z \nonumber\\
    &\quad + v_{12}\big[(1-\cos k)\,\sigma_x - \sin k\,\sigma_y\big]\otimes \tau_x .
\end{align}
Thus, 
\begin{equation}
    \mathcal{H}^a_k = (\mathbf{A}\cdot\boldsymbol\sigma)\otimes I 
    + (\mathbf{B}\cdot\boldsymbol\sigma)\otimes \tau_z
    + (\mathbf{C}\cdot\boldsymbol\sigma)\otimes \tau_x ,
\end{equation}
where, 
%$\mathbf{A}(k) = (t_{cf1}+t_{cf2}\cos k,\;\; t_{cf2}\sin k),\, \mathbf{B}(k) = (t_{cf1}'+t_{cf2}'\cos k,\;\; t_{cf2}'\sin k)$ and $\mathbf{C}(k) = \big(v_{12}(1-\cos k),\;\; -v_{12}\sin k\big).$
\begin{align}
    \mathbf{A}(k) &= (t_{cf1}+t_{cf2}\cos k,\;\; t_{cf2}\sin k), \\
    \mathbf{B}(k) &= (t_{cf1}'+t_{cf2}'\cos k,\;\; t_{cf2}'\sin k), \\
    \mathbf{C}(k) &= \big(v_{12}(1-\cos k),\;\; -v_{12}\sin k\big).
\end{align}

We can now perform a rotation in the $\tau$-space, with the unitary operator,  $U_\tau = e^{-i\frac{\alpha}{2}\tau_y}$, and, $U_{\text{full}} = I_\sigma \otimes U_\tau$, to obtain, 
\begin{align}
    \mathcal{H}^s_k &= (\mathbf{A}\cdot\boldsymbol\sigma)\otimes I +
\big[(\cos\alpha\,\mathbf{B}+\sin\alpha\,\mathbf{C})\cdot\boldsymbol\sigma\big]\otimes \tau_z \nonumber\\
    &\quad + \big[(-\sin\alpha\,\mathbf{B}+\cos\alpha\,\mathbf{C})\cdot\boldsymbol\sigma\big]\otimes \tau_x,
\end{align}
We must eliminate the \(\tau_x\) block to achieve a block-diagonal form for the Hamiltonian. This can be accomplished by imposing the condition

\begin{equation}
    -\sin\alpha\,\mathbf{B}(k) + \cos\alpha\,\mathbf{C}(k) \equiv \mathbf{0}, 
    \quad \forall\,k.
\end{equation}

We obtain two critical conditions by splitting this equation component-wise and solving the resulting expressions. From one of these components, we derive the rotation-angle condition:

\begin{equation}
    \tan\alpha=-\frac{v_{12}}{t'_{cf2}}=-\frac{2v_1}{t_{c_2}+t_{f_2}}, \, (v_{1},\, t_{c_2}+t_{f_2} \neq 0).    
\end{equation}

Utilizing this relation, we can further solve the other component, which provides a constraint on the parameter space necessary for the block diagonalization of the Hamiltonian. It is given by $t'_{cf1} = -t'_{cf2}$.

Reexpressing this relation using the parameters of the main Hamiltonian leads to the important condition:

\begin{equation}
    t_{c_1}+t_{c_2}+t_{f_1}+t_{f_2}=0.
\end{equation}

Under these conditions, the block diagonalized Hamiltonian takes the form,
\begin{equation}
    \mathcal{H}^s_k = (\mathbf{A}\cdot\boldsymbol\sigma)\otimes I
    + \frac{1}{\cos\alpha}\,(\mathbf{B}\cdot\boldsymbol\sigma)\otimes \tau_z.
    \label{Eq: rotated_hamiltonian}
\end{equation}

Equivalently, defining $W_0 = t_{cf1}'/\cos\alpha$ and $W_1 = t_{cf2}'/\cos\alpha$,
\begin{align}
    \mathcal{H}^s_k &= 
    (t_{cf1}+t_{cf2}\cos k)\,\sigma_x \otimes I +
    t_{cf2}\sin k\,\sigma_y \otimes I 
    \nonumber\\
    &\quad + 
    (W_0 + W_1\cos k)\,\sigma_x \otimes \tau_z + 
    W_1\sin k\,\sigma_y \otimes \tau_z .
\end{align}
This further implies the existence of an operator, defined as, 
$P = I_\sigma \otimes \tau_z$, such that, $[\mathcal{H}^s_k,\, P] = 0$, and the block diagonalized form may be written as, 
\begin{align*}
    \mathcal{H}^s_k=
    \begin{pmatrix}
           H_L(k) & 0\\
           0 & H_R(k),
    \end{pmatrix}
\end{align*}
where, \begin{align}
    H_L(k) = (\mathbf{A}+\tfrac{1}{\cos\alpha}\mathbf{B})\cdot\boldsymbol\sigma
\end{align} and,  \begin{align}
    H_R(k) = (\mathbf{A}-\tfrac{1}{\cos\alpha}\mathbf{B})\cdot\boldsymbol\sigma
\end{align} with, $\tan\alpha = -\frac{2v_{1}}{t_{c_2}+t_{f_2}}$.

Utilising the definitions of $\mathbf{A}$ and $\mathbf{B}$ each block can be written in the SSH form
\[
H_{\chi}(k) = \big(u_\chi+v_\chi\cos k\big)\sigma_x + v_\chi\sin k\,\sigma_y,\quad \chi=L,R,
\]
with effective parameters (sign \(s\) incorporated)
\begin{align}
    u_L &= t_{cf1} - sR, &\qquad v_L &= t_{cf2} + sR, \\
    u_R &= t_{cf1} + sR, &\qquad v_R &= t_{cf2} - sR.
\end{align}

Here, we define \( R = \sqrt{(t'_{cf2})^2 + v_{12}^2} \) as a nonnegative quantity, while \( s = \operatorname{sgn}(t'_{cf2}) \) retains the original sign of \( t'_{cf2} \). This formulation ensures that \( R \) remains positive and that the effective couplings carry the correct algebraic signs. According to the SSH topological criterion, block \( \chi \) is in the nontrivial phase when \( |u_\chi| < |v_\chi| \). Given that \( u_\chi \) and \( v_\chi \) are dependent on \( sR \), the sign of \( t^{'}_{cf2} \) influences the parameter space regions in which each block exhibits topological properties.

%%%%%%%%%%%%%%%%%%Number Density%%%%%%%%%%%%%%
%\clearpage
\section{Number Density}
\label{app:number_density}

\begin{figure}[htbp]
    \centering
    \includegraphics[width=1\linewidth]{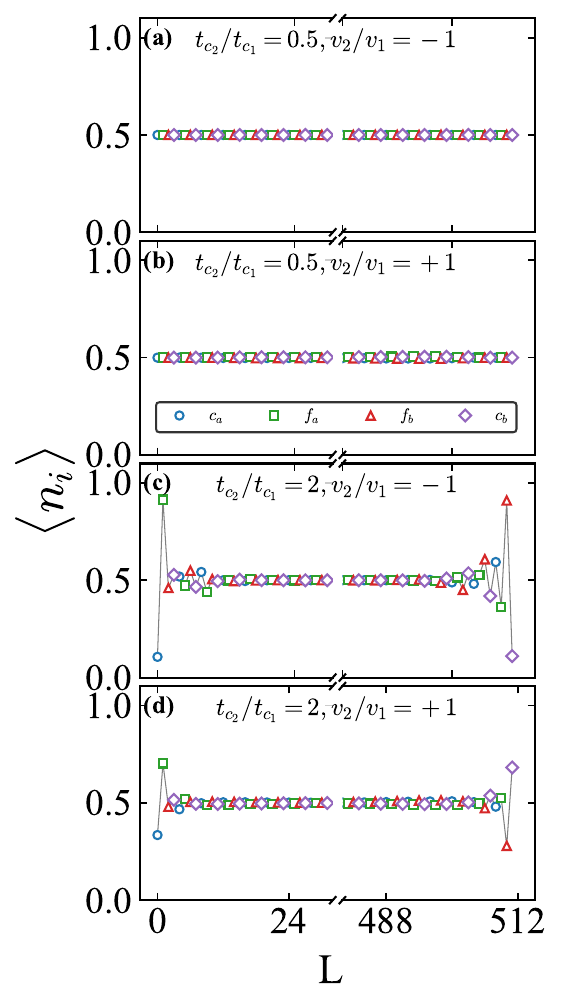}
    \caption{In this figure, we plot the number density $\langle n_i\rangle$ for interaction strength $V_{NN}=2$ and different parameter combinations (a) $t_{c_2}/t_{c_1}= 0.5,\, v_2/v_1=-1.0$ (trivial), (b) $t_{c_2}/t_{c_1}= 0.5,\, v_2/v_1=+1.0$ (trivial), (c) $t_{c_2}/t_{c_1}= 2.0,\, v_2/v_1=-1.0$ (topological), (d) $t_{c_2}/t_{c_1}= 2.0,\, v_2/v_1=+1$ (topological). 
    The topological phase hosts an edge imbalance. All other parameters are the same for all cases. System size considered for this calculation is $L= 512$. Markers with different shapes and colours represent a lattice's different sites ($c_a, c_b, f_a$ and $f_b$). 
    }
    \label{fig:number_density}
\end{figure}

Figure~\ref{fig:number_density}(a-d) displays the number density plots corresponding to a defined set of parameters that fall within the parameter region of interacting phase A${}^{'}$ (panels a and b) and phase C${}^{'}$ (panels c and d) for interaction strength $ V_{NN} = 2 $. 
%\clearpage
The presence of edge imbalance in the number density plots of panels c and d corroborates our findings in Section~\ref{sec:results_int_S_edge}, indicating that phase C${}^{'}$ is topologically non-trivial. Conversely, the lack of edge imbalance in panels a and b supports the classification of phase A${}^{'}$ as trivial.

\section{Central Charge}
\label{app:central_charge}

The bipartition entanglement entropy \( S(l) \) of a subsystem of length \( l \) is described in Equation~\ref{cc} for the ground state of the system. For interacting systems, we compute \( S(l) \) using the matrix product state (MPS) obtained through the density matrix renormalization group (DMRG) method for various lengths \( l \).
\begin{figure}[htbp]
    \centering
    \includegraphics[width=1\linewidth]{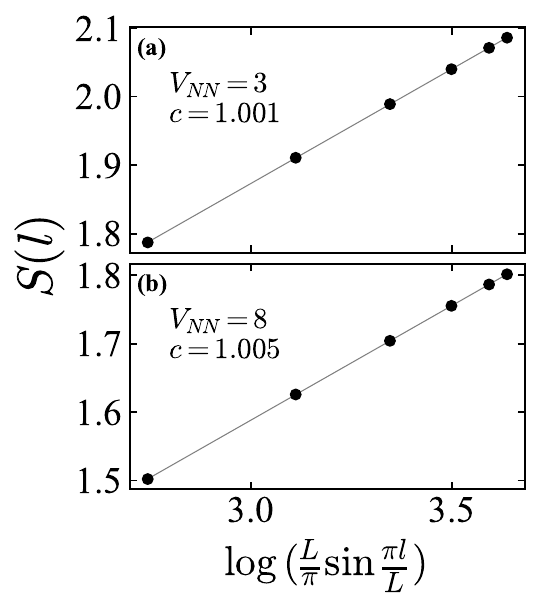}
    \caption{This figure depicts a scatter plot of \( S(l) \) versus \( \log\left(\frac{L}{\pi} \sin{\frac{\pi l}{L}}\right) \), along with a linear fit. Top and bottom panels are for interaction \( V_{NN} = 3 \) and \( V_{NN} = 8 \), respectively. Central charge $c$ shown in the panels (a) (1.001) and (b) (1.005) is calculated from the slope of the fitted line.}
    \label{fig:c_fit_few}
\end{figure}
 With periodic boundary conditions, the slope of the \( S(l) \) versus \( \log\left(\frac{L}{\pi}\sin{\frac{\pi l}{L}}\right) \) plot yields \( \frac{c}{3} \), allowing us to derive \( c \). Values of \( c \) are calculated for different interaction strengths \( V_{NN} \). Figure~\ref{fig:c_fit_few}(a,b) displays scatter plots (blue circles) with linear fits for \( V_{NN} = 8 \) and \( V_{NN} = 3 \),  respectively.

\begin{greenblock}
    \section{Many-body chiral symmetry and the
\texorpdfstring{$\mathbb Z_4$}{Z4} edge index}
\label{app:manybody_spt}
\label{app:chiral_interaction}% compatibility alias for the baseline reference
The interacting classification is formulated using the antiunitary many-body chiral symmetry $\mathcal S$ defined in Eq.~\eqref{eq:manybody_chiral_action}. The free Bloch Hamiltonian is simultaneously organized by the single-particle sublattice operator $\sigma_z\otimes\mathbb I$, which produces its block-off-diagonal structure and integer winding number. These two symmetry actions play complementary roles: $\mathcal S$ defines the interacting edge algebra, while the sublattice operator organizes the free band problem.

\subsection{Edge representation}

Consider $q$ elementary complex edge modes $d_m$ with
\begin{equation}
\mathcal S_e d_m\mathcal S_e^{-1}=d_m^\dagger,
\qquad
m=1,\ldots,q.
\label{eq:edge_S_action}
\end{equation}
Define
\begin{equation}
P_e=(-1)^{\sum_m d_m^\dagger d_m},
\,
\gamma_m=d_m+d_m^\dagger,
\,
X_q=\gamma_1\gamma_2\cdots\gamma_q.
\end{equation}
Let $K_e$ denote complex conjugation in the occupation-number basis,
chosen so that the matrices $d_m$ and $\gamma_m$ are real. Since
$\gamma_n$ anticommutes with $d_m$ for $n\neq m$,
\begin{equation}
X_qd_mX_q^{-1}
=
(-1)^{q-1}d_m^\dagger.
\label{eq:Xq_action}
\end{equation}
A representative of the antiunitary edge symmetry is therefore
\begin{equation}
\mathcal S_e
=
\begin{cases}
X_qK_e,&q\ {\rm odd},\\[2mm]
P_eX_qK_e,&q\ {\rm even}.
\end{cases}
\label{eq:Se_representation}
\end{equation}
For odd $q$, $X_q$ already implements
$d_m\mapsto d_m^\dagger$; for even $q$, the additional parity operator
removes the minus sign in Eq.~\eqref{eq:Xq_action}.

Using
\begin{equation}
P_eX_qP_e^{-1}=(-1)^qX_q,
\qquad
X_q^2=(-1)^{q(q-1)/2},
\end{equation}
and the reality of $P_e$ and $X_q$, one obtains
\begin{equation}
P_e\mathcal S_eP_e^{-1}=(-1)^q\mathcal S_e,
\qquad
\mathcal S_e^2=(-1)^{q(q-1)/2}.
\label{eq:edge_projective_signs}
\end{equation}
The four inequivalent sign pairs are
\begin{equation}
\begin{array}{c|c|c}
q\pmod4&
P_e\mathcal S_eP_e^{-1}/\mathcal S_e&
\mathcal S_e^2
\\ \hline
0&+1&+1\\
1&-1&+1\\
2&+1&-1\\
3&-1&-1
\end{array}.
\end{equation}
The pattern repeats at $q+4$. The completeness of these four classes is
the charge-conserving chiral $\mathbb Z_4$ classification
\cite{LeeShiozakiHsieh2024,BourneOgata2021}.

The triviality of four copies may also be seen locally. For $q=4$,
$\mathcal S_e$ is parity even and squares to $+1$. In a consistent
occupation ordering, the charge-neutral state
\begin{equation}
\lvert\Omega_4\rangle
=
\frac{1}{\sqrt2}
\left(
\lvert1100\rangle-\lvert0011\rangle
\right)
\end{equation}
may be chosen $\mathcal S_e$ invariant. The local edge Hamiltonian
\begin{equation}
H_{4,e}
=
UQ_e^2
+
\Delta
\left(
\mathbb I-
\lvert\Omega_4\rangle\langle\Omega_4\rvert
\right),
\,
Q_e=\sum_{m=1}^4
\left(d_m^\dagger d_m-\frac12\right),
\end{equation}
is charge conserving and $\mathcal S_e$ symmetric and has a unique
ground state. Thus four elementary edge generators can be removed by a
local symmetry-preserving interaction.

\subsection{Connection to the ladder}

In the block-diagonal limit of Appendix~\ref{app:decoupled_SSH}, a
topological SSH block $\chi=L,R$ carries the left-edge mode
\begin{equation}
d_{\chi,L}
=
\mathcal N_\chi
\sum_{j\geq1}
\left(
-\frac{u_\chi}{v_\chi}
\right)^{j-1}
a_{\chi,j},
\qquad
|u_\chi|<|v_\chi|.
\label{eq:SSH_edge_mode}
\end{equation}
The microscopic many-body chiral symmetry gives
\begin{equation}
\mathcal S d_{\chi,L}\mathcal S^{-1}
=
d_{\chi,L}^\dagger,
\end{equation}
up to an unimportant basis phase. Each topological SSH block therefore
contributes one elementary edge generator. The representative phases
contain
\begin{equation}
q_A=0,\qquad q_B=1,\qquad q_C=2.
\end{equation}
Since the manuscript uses the opposite orientation for the winding,
$q=-\nu$, and
\begin{equation}
I_{\mathcal S}=\nu\pmod4.
\end{equation}
This gives Eq.~\eqref{eq:interacting_Z4_labels} after
symmetry-preserving adiabatic continuation into the three interacting
gapped regions.

\section{Continuum projection at the
\texorpdfstring{$A$-$C$}{A-C} boundary}
\label{app:AC_projection}

Here we give the details of the low-energy projection used in
Sec.~\ref{sec:pin_shift_mechanism}. On the free $A$-$C$ line,
$t_{f_1}=t_{f_2}\equiv t_f$ and
$t_{c_1}=t_{c_2}\equiv\bar t_c$. Let
\begin{equation}
s_0=\sin\frac{k_0}{2},
\qquad
c_0=\cos\frac{k_0}{2},
\end{equation}
where $k_0$ obeys Eq.~\eqref{eq:AC_closing_condition}.
At valley $k_\lambda=\lambda k_0$, $\lambda=\pm$, normalized null
vectors of $h^\dagger(k_\lambda)$ and $h(k_\lambda)$ may be chosen as
\begin{align}
\widehat r_\lambda
&=
\frac{1}{\mathcal N_a}
\begin{pmatrix}
i\lambda v_1s_0\\
\bar t_c c_0
\end{pmatrix},
&
\mathcal N_a^2
&=
v_1^2s_0^2+\bar t_c^{\,2}c_0^2,
\nonumber\\
\widehat\ell_\lambda
&=
\frac{1}{\mathcal N_b}
\begin{pmatrix}
i\lambda v_2s_0\\
\bar t_c c_0
\end{pmatrix},
&
\mathcal N_b^2
&=
v_2^2s_0^2+\bar t_c^{\,2}c_0^2.
\label{eq:AC_null_vectors}
\end{align}
The corresponding light states of the four-component Bloch
Hamiltonian are
\begin{equation}
\Phi_{\lambda,1}
=
\begin{pmatrix}
\widehat r_\lambda\\0
\end{pmatrix},
\qquad
\Phi_{\lambda,2}
=
\begin{pmatrix}
0\\\widehat\ell_\lambda
\end{pmatrix}.
\end{equation}
The remaining singular value, denoted by $\Lambda_0$, stays finite.
Projecting the momentum deviation and $\delta t_c$ onto this subspace
gives, after a constant Pauli-space rotation,
Eq.~\eqref{eq:HAC_Dirac_main}, with
\begin{equation}
v_*
=
\frac{
2|\sin k_0|\,|v_1v_2-\bar t_c t_f|
}{\Lambda_0},
\qquad
m_*
=
\frac{2t_f\sin k_0}{\Lambda_0}\,\delta t_c .
\label{eq:vstar_mstar_main}
\end{equation}
Thus $\delta t_c$ is the common mass of the two valley Dirac theories.

\subsection{Projected interaction}

Let $a_\lambda$ and $b_\lambda$ denote the light fields associated
with $\widehat r_\lambda$ and $\widehat\ell_\lambda$, respectively.
Keeping the nonoscillating terms, projection of
Eq.~\eqref{eqn:H_int1} gives
\begin{equation}
H_{\rm int}^{AC}
=
\int dx\,
\left[
g_a\rho_{a,+}\rho_{a,-}
+
g_b\rho_{b,+}\rho_{b,-}
\right],
\label{eq:AC_projected_density}
\end{equation}
where
\begin{align}
g_a
&=
\frac{
4V_{NN}v_1^2\bar t_c^{\,2}s_0^2c_0^2
}{
\left(
v_1^2s_0^2+\bar t_c^{\,2}c_0^2
\right)^2
},
\nonumber\\
g_b
&=
\frac{
4V_{NN}v_2^2\bar t_c^{\,2}s_0^2c_0^2
}{
\left(
v_2^2s_0^2+\bar t_c^{\,2}c_0^2
\right)^2
}.
\label{eq:AC_projected_couplings}
\end{align}
In particular, $g_a$ and $g_b$ have the sign of $V_{NN}$.

With
\begin{equation}
a_\lambda=\frac{R_\lambda+L_\lambda}{\sqrt2},
\qquad
b_\lambda=\frac{R_\lambda-L_\lambda}{\sqrt2},
\end{equation}
and
\begin{equation}
D_\lambda
=
R_\lambda^\dagger R_\lambda
+
L_\lambda^\dagger L_\lambda,
\,
B_\lambda=R_\lambda^\dagger L_\lambda,
\,
M_\lambda=B_\lambda+B_\lambda^\dagger .
\end{equation}
the interaction becomes
\begin{align}
H_{\rm int}^{AC}
&=
\frac14\int dx\,
\Big[
(g_a+g_b)
\left(
D_+D_-+M_+M_-
\right)
\nonumber\\
&\hspace{2.0cm}
+
(g_a-g_b)
\left(
D_+M_-+M_+D_-
\right)
\Big].
\label{eq:AC_projected_RL}
\end{align}
The first mass product contains
\begin{align}
M_+M_-
&=
\mathcal O_t+\mathcal O_r,
\nonumber\\
\mathcal O_t
&=
B_+B_-+B_+^\dagger B_-^\dagger,
\nonumber\\
\mathcal O_r
&=
B_+B_-^\dagger+B_+^\dagger B_- ,
\end{align}
which establishes that the projected density interaction generates both intervalley channels; their bosonized forms are given in Eq.~\eqref{eq:AC_two_cosines} below.

\subsection{Weak-coupling continuum interpretation}

To illustrate the effect of the intervalley terms, consider first the form-factor-symmetric limit $g_a=g_b$, for which the mixed terms in Eq.~\eqref{eq:AC_projected_RL} vanish. Using the bosonization convention
\cite{Giamarchi2004}
\begin{equation}
B_\lambda
=
R_\lambda^\dagger L_\lambda
\sim
\frac{\kappa_\lambda}{2\pi a_0}
e^{2i\phi_\lambda},
\qquad
\phi_{t,r}
=
\frac{\phi_+\pm\phi_-}{\sqrt2},
\end{equation}
the two intervalley operators become
\begin{equation}
\mathcal O_t\propto\cos(2\sqrt2\,\phi_t),
\qquad
\mathcal O_r\propto\cos(2\sqrt2\,\phi_r).\label{eq:AC_two_cosines}
\end{equation}
The Gaussian part of each collective sector may be written as
\begin{equation}
H_\mu
=
\frac{u_\mu}{2\pi}
\int dx
\left[
K_\mu(\partial_x\theta_\mu)^2
+
\frac{1}{K_\mu}(\partial_x\phi_\mu)^2
\right],
\qquad
\mu=t,r,
\end{equation}
which defines the corresponding Luttinger parameters $K_t$ and $K_r$. At the free Dirac point $K_t=K_r=1$.

The forward-scattering part of a repulsive density interaction shifts the two collective stiffnesses in opposite directions,
\begin{equation}
K_t
=
1-\kappa V_{NN}+O(V_{NN}^2),
\,
K_r
=
1+\kappa V_{NN}+O(V_{NN}^2),
\,
\kappa>0 .
\end{equation}
Since the two cosine operators have scaling dimensions $2K_t$ and $2K_r$, respectively, this symmetric limit provides a weak-coupling realization in which one collective perturbation becomes relevant while the other becomes irrelevant. The resulting infrared structure contains one massive and one gapless collective sector and therefore has $c=1$.

For generic microscopic parameters $g_a\neq g_b$, additional marginal operators in Eq.~\eqref{eq:AC_projected_RL} mix the two collective channels. The DMRG central-charge calculation establishes the $c=1$ critical behaviour on the parameter cuts studied in the main text; the analysis above provides the corresponding weak-coupling continuum mechanism in the analytically transparent symmetric limit.

\section{Second-order shift of the
\texorpdfstring{$B$-$C$}{B-C} boundary}
\label{app:BC_self_energy}

We derive here the weak-coupling displacement of the $B$-$C$
boundary used in Sec.~\ref{sec:BC_shift_main}. Throughout this
appendix the orbital basis is
\begin{equation}
\Psi_k=
\begin{pmatrix}
c_{a,k}&f_{a,k}&c_{b,k}&f_{b,k}
\end{pmatrix}^{T},
\end{equation}
so that the indices $1,2,3,4$ denote
$c_a,f_a,c_b,f_b$, respectively.

\subsection{Light subspace at the free transition}

On the free $B$-$C$ boundary we set
\begin{equation}
v_2=0,
\qquad
t_{f_1}=t_{f_2}\equiv t_f,
\end{equation}
and expand around $k=\pi$. Defining
\begin{equation}
m_c=t_{c_1}-t_{c_2},
\qquad
M=\sqrt{m_c^2+4v_1^2},
\end{equation}
the chiral block at the gap-closing momentum becomes
\begin{equation}
h_0(\pi)
=
\begin{pmatrix}
m_c&0\\
2v_1&0
\end{pmatrix}.
\label{eq:BC_h_pi}
\end{equation}
Hence the full four-component Hamiltonian has two zero modes,
which may be chosen as
\begin{equation}
u_d
=
\frac{1}{M}
\begin{pmatrix}
2v_1\\
-m_c\\
0\\
0
\end{pmatrix},
\qquad
u_e
=
\begin{pmatrix}
0\\
0\\
0\\
1
\end{pmatrix}.
\label{eq:BC_zero_vectors}
\end{equation}
These correspond to the real-space light fields
\begin{equation}
d_j=
\frac{2v_1c_{a,j}-m_cf_{a,j}}{M},
\qquad
e_j=f_{b,j},
\end{equation}
used in Eq.~\eqref{eq:BC_light_fields}.

Let
\begin{equation}
P_L=
\begin{pmatrix}
u_d&u_e
\end{pmatrix}
\end{equation}
be the projector onto this two-dimensional light subspace. The full
inverse Green function is
\begin{equation}
K(i\omega,k)
\equiv
G^{-1}(i\omega,k)
=
G_0^{-1}(i\omega,k)-\Sigma(i\omega,k),
\end{equation}
and its light-sector projection is
\begin{equation}
K_{LL}=P_L^\dagger K P_L .
\label{eq:BC_projector_definition}
\end{equation}

At $k=\pi$, the bare $v_2$ term satisfies
\begin{equation}
[\mathcal H(\pi)]_{14}=-2v_2,
\end{equation}
while $[\mathcal H(\pi)]_{24}=0$ for
$t_{f_1}=t_{f_2}$. Therefore
\begin{align}
[K_{LL}(0,\pi)]_{12}
&=
\frac{2v_1}{M}K_{14}(0,\pi)
-\frac{m_c}{M}K_{24}(0,\pi)
\nonumber\\
&=
\frac{2v_1}{M}
\left[
2v_2-\Sigma_{14}(0,\pi)
\right]
+
\frac{m_c}{M}\Sigma_{24}(0,\pi).
\label{eq:BC_projection_exact}
\end{align}
Thus $\Sigma_{14}$ describes the induced
$c_a^\dagger f_b$ channel and $\Sigma_{24}$ the induced
$f_a^\dagger f_b$ channel; both project onto the same
$d^\dagger e+\mathrm{H.c.}$ Dirac mass.

A single density vertex acts entirely within either the
$a$-sublattice pair $(c_a,f_a)$ or the $b$-sublattice pair
$(c_b,f_b)$. Consequently a first-order contraction cannot connect
the $d$ and $e$ light sectors:
\begin{equation}
\Sigma_{14}^{(1)}(0,\pi)
=
\Sigma_{24}^{(1)}(0,\pi)
=
0.
\label{eq:BC_first_order_zero}
\end{equation}
The leading off-sublattice mass correction therefore requires one
density vertex on each sublattice and starts at $O(V_{NN}^2)$.

\subsection{Bare lattice Green function}

At $v_2=0$, write the off-diagonal block as
\begin{equation}
h_0(k)
=
\begin{pmatrix}
a_k&0\\
b_k&c_k
\end{pmatrix},
\label{eq:BC_h0_triangular}
\end{equation}
with
\begin{equation}
a_k=t_{c_1}+t_{c_2}e^{-ik},
\,
b_k=v_1(1-e^{-ik}),
\,
c_k=-(t_{f_1}+t_{f_2}e^{-ik}).
\label{eq:BC_abc}
\end{equation}
For
\begin{equation}
G^{(0)}(i\omega,k)
=
\left[
i\omega-\mathcal H_0(k)
\right]^{-1},
\end{equation}
the common denominator of the matrix elements required below is
\begin{equation}
\Delta_k(\omega)
=
\omega^4
+
\omega^2
\left(
|a_k|^2+|b_k|^2+|c_k|^2
\right)
+
|a_k|^2|c_k|^2 .
\label{eq:BC_G_denominator}
\end{equation}
Direct inversion gives
\begin{align}
G_{13}^{(0)}(i\omega,k)
&=
-\frac{
a_k(\omega^2+|c_k|^2)
}{
\Delta_k(\omega)
},
&
G_{24}^{(0)}(i\omega,k)
&=
-\frac{
c_k(\omega^2+|a_k|^2)
}{
\Delta_k(\omega)
},
\nonumber\\
G_{32}^{(0)}(i\omega,k)
&=
-\frac{
b_k^*\omega^2
}{
\Delta_k(\omega)
},
&
G_{31}^{(0)}(i\omega,k)
&=
-\frac{
a_k^*(\omega^2+|c_k|^2)
}{
\Delta_k(\omega)
},
\nonumber\\
G_{23}^{(0)}(i\omega,k)
&=
-\frac{
b_k\omega^2
}{
\Delta_k(\omega)
},
&
G_{14}^{(0)}(i\omega,k)
&=
\frac{
a_kb_k^*c_k
}{
\Delta_k(\omega)
}.
\label{eq:BC_G_elements}
\end{align}

\subsection{Second-order self-energy}

For the off-sublattice mass channels, the relevant quartic part of the
Euclidean interaction may be written as
\begin{equation}
S_{\rm int}
=
V_{NN}\int_x
\left(
A_x+B_x
\right),
\end{equation}
where
\begin{equation}
A_x=
\bar c_a c_a\,\bar f_a f_a,
\qquad
B_x=
\bar c_b c_b\,\bar f_b f_b,
\qquad
\int_x\equiv\int d\tau\sum_j .
\end{equation}
The centering of the densities affects the tadpole and chemical-potential
pieces but leaves these quartic vertices unchanged.

The connected second-order contribution to the one-particle-irreducible
effective action is
\begin{equation}
\delta S_{\rm eff}^{(2)}
=
-\frac{V_{NN}^2}{2}
\int_{x,y}
\left\langle
(A_x+B_x)(A_y+B_y)
\right\rangle_{0,c}.
\label{eq:BC_second_order_action}
\end{equation}
Only the cross terms $A_xB_y$ and $B_xA_y$ connect opposite
sublattices. After exchanging $x\leftrightarrow y$, they give the same
translation-invariant contribution and cancel the explicit factor
$1/2$ in Eq.~\eqref{eq:BC_second_order_action}.

For the $(1,4)$ channel, leave the external bilinear
$\bar c_a(x)f_b(y)$ uncontracted. The remaining fields have the
connected contraction
\begin{equation}
c_a(x)\leftrightarrow\bar c_b(y),
\qquad
c_b(y)\leftrightarrow\bar f_a(x),
\qquad
f_a(x)\leftrightarrow\bar f_b(y),
\end{equation}
which gives
\begin{equation}
\Sigma_{14}^{(2)}(x-y)
=
-V_{NN}^2
G_{13}^{(0)}(x-y)
G_{32}^{(0)}(y-x)
G_{24}^{(0)}(x-y).
\label{eq:Sigma14_coordinate}
\end{equation}

For the $(2,4)$ channel, leaving
$\bar f_a(x)f_b(y)$ external gives direct and exchange contractions,
\begin{align}
\Sigma_{24}^{(2)}(x-y)
&=
-V_{NN}^2
\Big[
G_{13}^{(0)}(x-y)
G_{31}^{(0)}(y-x)
G_{24}^{(0)}(x-y)
\nonumber\\
&\hspace{1.5cm}
-
G_{14}^{(0)}(x-y)
G_{23}^{(0)}(x-y)
G_{31}^{(0)}(y-x)
\Big].
\label{eq:Sigma24_coordinate}
\end{align}

To Fourier transform these expressions, define
\begin{equation}
\int_{\Omega,\nu,q,p}
\equiv
\int_{-\infty}^{\infty}\frac{d\Omega}{2\pi}
\int_{-\infty}^{\infty}\frac{d\nu}{2\pi}
\int_{-\pi}^{\pi}\frac{dq}{2\pi}
\int_{-\pi}^{\pi}\frac{dp}{2\pi},
\label{eq:BC_integral_measure}
\end{equation}
and, for external momentum $k=\pi$,
\begin{equation}
Q=q+p-\pi
\qquad
(\mathrm{mod}\ 2\pi).
\end{equation}
Equations~\eqref{eq:Sigma14_coordinate} and
\eqref{eq:Sigma24_coordinate} then become
\begin{align}
\Sigma_{14}^{(2)}(0,\pi)
&=
-V_{NN}^2
\int_{\Omega,\nu,q,p}
G_{13}^{(0)}(i\Omega,q)
G_{24}^{(0)}(i\nu,p)
\nonumber\\
&\hspace{2cm}\times
G_{32}^{(0)}
\bigl(i(\Omega+\nu),Q\bigr),
\label{eq:Sigma14_integral}
\end{align}
and
\begin{align}
\Sigma_{24}^{(2)}(0,\pi)
&=
-V_{NN}^2
\int_{\Omega,\nu,q,p}
\Big[
G_{13}^{(0)}(i\Omega,q)
G_{24}^{(0)}(i\nu,p)
\nonumber\\
&\hspace{1.5cm}\times
G_{31}^{(0)}
\bigl(i(\Omega+\nu),Q\bigr)
\nonumber\\
&\hspace{0.6cm}
-
G_{14}^{(0)}(i\Omega,q)
G_{23}^{(0)}(i\nu,p)
G_{31}^{(0)}
\bigl(i(\Omega+\nu),Q\bigr)
\Big].
\label{eq:Sigma24_integral}
\end{align}
These equations give the microscopic second-order lattice
self-energies entering the light Dirac mass.

\subsection{Shifted critical condition}

Substituting
\begin{equation}
\Sigma_{14}
=
\Sigma_{14}^{(2)}+O(V_{NN}^3),
\qquad
\Sigma_{24}
=
\Sigma_{24}^{(2)}+O(V_{NN}^3)
\end{equation}
into Eq.~\eqref{eq:BC_projection_exact} gives
\begin{align}
[K_{LL}(0,\pi)]_{12}
&=
\frac{4v_1}{M}v_2
-\frac{2v_1}{M}\Sigma_{14}^{(2)}(0,\pi)
\nonumber\\
&\quad+
\frac{m_c}{M}\Sigma_{24}^{(2)}(0,\pi)
+
O(V_{NN}^3).
\label{eq:BC_KLL_shift}
\end{align}
The $B$-$C$ transition is obtained by setting the light Dirac mass to
zero, $[K_{LL}(0,\pi)]_{12}=0$. Hence
\begin{equation}
\boxed{
2v_{2,c}(V_{NN})
=
\Sigma_{14}^{(2)}(0,\pi)
-
\frac{m_c}{2v_1}
\Sigma_{24}^{(2)}(0,\pi)
+
O(V_{NN}^3)
}.
\end{equation}
This reproduces Eq.~\eqref{eq:v2_shift_main}. Since the two self-energies begin at second order, the leading displacement of the critical line is $O(V_{NN}^2)$.

For the representative cut
\begin{equation}
t_{c_1}=1,\qquad
t_{c_2}=2,\qquad
t_{f_1}=t_{f_2}=0.3,\qquad
v_1=-0.3,
\end{equation}
numerical evaluation of the lattice integrals in
Eqs.~\eqref{eq:Sigma14_integral} and
\eqref{eq:Sigma24_integral} gives
\begin{equation}
\Sigma_{14}^{(2)}(0,\pi)
\simeq
6.9\times10^{-3}V_{NN}^2,
\qquad
\Sigma_{24}^{(2)}(0,\pi)
\simeq
2.69\times10^{-2}V_{NN}^2 .
\label{eq:BC_self_energy_numeric}
\end{equation}
For $m_c=t_{c_1}-t_{c_2}=-1$, the resulting weak-coupling shift is
\begin{equation}
v_{2,c}
\simeq
-1.90\times10^{-2}V_{NN}^2,
\qquad
\frac{v_{2,c}}{v_1}
\simeq
6.3\times10^{-2}V_{NN}^2 .
\label{eq:BC_shift_numeric}
\end{equation}
The positive shift in $v_{2,c}/v_1$ has the same orientation as the
motion of the gap minimum in Fig.~\ref{fig:gap_analysis_full}(c).
Thus the lattice self-energy calculation reproduces both the quadratic weak-coupling onset and the direction of the interacting boundary
displacement.
\end{greenblock}

%\clearpage
\end{appendix}
\bibliography{References}

\section*{Table of Contents}
\noindent\begin{minipage}{\linewidth}
\centering
\includegraphics[width=55mm]{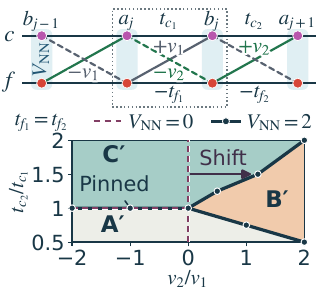}\par\medskip
\raggedright
At half filling, charge conservation and an antiunitary many-body chiral symmetry protect the interacting phases. Together with reflection symmetry, this structure pins the numerically observed $A'$–$C'$ transition boundary, even as gapless modes fall from two to one. Both the $B'$–$C'$ and $B'$–$A'$ transitions retain a single gapless mode, but the boundary is renormalized, and the shifting of the renormalized boundary is proportional to the square of the interaction strength.
\end{minipage}\par

\end{document}